\documentclass[
   aps,pre,twocolumn,
   reprint,
   superscriptaddress,
   nofootinbib,
   longbibliography,
   floatfix,
   amssymb,
]{revtex4-2}
 
\usepackage{graphicx} 
\usepackage{amsmath}
\usepackage{xcolor}
  
\date{\today}
 
\usepackage{enumitem}
\usepackage[normalem]{ulem}   

\definecolor{linkColor}{rgb}{0,0.3,0.7}
\usepackage{hyperref}
\hypersetup{colorlinks=true,
           allcolors=linkColor,
           pdfborder={0 0 0},
           pdfencoding=auto
 }
 
\begin{document}

\title{Stochastic Yield Catastrophe in Delay-Facilitated Self-Assembly}
\author{Richard Swiderski}
\thanks{R.S. and S.A. contributed equally.}
\author{Severin Angerpointner}
\thanks{R.S. and S.A. contributed equally.}
\affiliation{Arnold Sommerfeld Center for Theoretical Physics and Center for NanoScience, Department of Physics, Ludwig-Maximilians-Universit\"at M\"unchen, Theresienstra\ss e 37, D-80333 Munich, Germany}

\author{Erwin Frey}
\email[Corresponding author: ]{frey@lmu.de}
\affiliation{Arnold Sommerfeld Center for Theoretical Physics and Center for NanoScience, Department of Physics, Ludwig-Maximilians-Universit\"at M\"unchen, Theresienstra\ss e 37, D-80333 Munich, Germany}
\affiliation{Max Planck School Matter to Life, Hofgartenstraße 8, 80539 Munich, Germany}

\begin{abstract}
Self-assembly of supramolecular structures in cells and synthetic applications often proceeds under unfavorable biochemical conditions and at low copy numbers of final target structures, ranging from tens of bacterial microcompartments to a single bacterial flagellum per cell.
Spatial organization through coupled reaction compartments of different reactivity (delay-facilitated assembly) can recover high yield in such environments at the mean-field level, but its robustness to stochastic fluctuations at low target numbers is unclear.
Using stochastic simulations of a minimal two-compartment model, we show that delay-facilitated assembly is susceptible to a stochastic yield catastrophe at low target numbers: even when each compartment in isolation allows for high-yield assembly, slow exchange between them induces a substantial drop in the final yield.
We trace the mechanism to a specific assembly stage, where the random order of rate-limiting exchange events of subunits and partially completed structures determines the ratio of productive growth to excess nucleation.
Restricting the exchange of larger structures---either by suppressing it entirely or letting exchange rates decrease with size---restores most of the yield without altering the mean-field behavior.
The same phenomenology appears for two-dimensional hexagonal subunits and in a cytosol-membrane geometry, where diffusion-limited exchange naturally implements the required size dependence.
Our results show that equal success of assembly strategies at high target numbers does not imply their equal success at low target numbers, and that competing slow events occurring in random order are a common signature of stochastic yield catastrophes.
\end{abstract}
 
\maketitle
 
\section{Introduction}
\label{sec:intro}
 
Self-assembly of supramolecular structures out of individual smaller subunits is a crucial process for living cells, synthetic biology, and nanotechnology.
In living systems, well-studied examples of such self-assembling structures are ribosomes \citep{Shajani2011RibosomeBacterial, Basler2018RibosomeAssemblyEukaryotes}, virus capsids \citep{Zlotnick2011VirusCapsids, Perlmutter2015VirusCapsids}, bacterial microcompartments \citep{Kerfeld2018BMCreview}, or flagella \citep{ChevanceFlagellumOverview, Nakamura2024}.
Inspired by such biological systems, it is possible to design completely new synthetic structures at the molecular scale, e.\,g., containers for virus trapping \citep{siglProgrammableIcosahedralShell2021, monferrerDNAOrigamiTraps2023}, or targeted molecule delivery \citep{khmelinskaia_structure-based_2021, jiangDNAOrigamiMolecular2024}.
To understand biological systems and improve synthetic applications, it is essential to uncover the underlying requirements for effective self-assembly.
One of the core requirements for successful assembly is a timescale separation between a slow nucleation process and a subsequent fast growth process to avoid kinetic traps, i.e., metastable states that require very long times to relax to the minimal free energy state of completed assemblies
\citep{zlotnickTheoreticalModelSuccessfully1999, morozov_assembly_2009, hagan_understanding_2010, haganMechanismsKineticTrapping2011, keThreeDimensionalStructuresSelfAssembled2012, weiComplexShapesSelfassembled2012, reinhardtNumericalEvidenceNucleated2014, gartnerStochasticYieldCatastrophes2020}.
 
Reaching such a separation of timescales for a given type of assembly and biochemical environment requires different control strategies to achieve high yield of the desired structures.
For instance, assembling multi-component DNA structures typically requires careful design of the molecular interactions \citep{rovigattiSimpleSolutionProblem2022, pintoDesignStrategiesSelfassembly2023}, or time-dependent temperature protocols \citep{jacobsRationalDesignSelfassembly2015}.
Other strategies involve spatial organization of the assembly process through phase separation of the medium surrounding the assembly \citep{haganSelfassemblyCoupledLiquidliquid2023, laha_chemical_2024} or by the subunits themselves \citep{bartolucci_interplay_2024}.
We previously showed that the yield in unfavorable chemical environments can be generally improved through delay-facilitated assembly, i.e., by limiting the overall nucleation rate via particle exchange between a high-reactivity and low-reactivity compartment \citep{angerpointnerDelayfacilitatedSelfassemblyCompartmentalized2025}.
 
For biological assemblies and nanoscale applications, the total number of desired target structures can vary over several orders of magnitude.
While ribosomes are usually assembled in $\mathcal{O}(10^4)$ copies per cell \citep{phillipsPhysicalBiologyCell2013odetoecoli}, different virus families are typically copied in a large range of $\mathcal{O}(10\text{--}10^4)$ times per cell \citep{delbruckBurstSizeDistribution1945, brussaardViralControlPhytoplankton2004, paradaViralBurstSize2006, chenDeterminationVirusBurst2007}.
Bacterial microcompartments, such as carboxysomes, or flagella assemble in even fewer numbers with only $\mathcal{O}(1\text{--}10)$ copies produced per cell \citep{hillLifeCycleCyanobacterial2020}, with many species only assembling a single flagellum \citep{Kojima2020, Bange2025}.
 
Across this large range of target numbers and the different types of assembly strategies, mean-field models describing the bulk behavior of large systems cannot always accurately predict the overall assembly yield and random fluctuations might need to be taken into account.
For instance, fluctuations can populate kinetic trap states not captured by deterministic rate equations~\citep{haganMechanismsKineticTrapping2011}, make the initial assembly kinetics depend on the system size~\citep{tiwariStochasticLagTime2016, michaelsFluctuationsKineticsLinear2016}, and render mean assembly times sensitive to the assumed distribution of initial states~\citep{davisInitialConditionStochastic2016}.
Crucially, fluctuations can also influence the final assembly yield.
Even at large target numbers, identical mixtures of different subunits can stochastically yield different compositions of final target structures \citep{schaefferStochasticEmergenceTwo2022}.
At low target numbers, certain assembly strategies for multi-component assembly in well-mixed systems have been shown to be susceptible to a ``stochastic yield catastrophe'', i.e., a drastic reduction in overall yield of finished structures due to fluctuations, whereas other strategies remain robust \citep{gartnerStochasticYieldCatastrophes2020}.
It is therefore generally unclear whether a given mean-field assembly strategy remains robust against fluctuations, and which model details govern this robustness.
In this work, we ask whether spatially organized self-assembly processes are susceptible to stochastic yield reductions at low target numbers and, if so, how detrimental effects of stochastic fluctuations can be mitigated.
 
To approach these questions, we build upon our previous study on delay-facilitated assembly~\citep{angerpointnerDelayfacilitatedSelfassemblyCompartmentalized2025} and extend it to the regime of low target numbers.
Although the impact of fluctuations is negligible in a single well-mixed compartment, we find that they can drastically reduce the overall yield in a two-compartment system.
Our analysis here shows that the stochastic exchange of low numbers of unfinished structures between the spatial compartments makes the final yield dependent on the random order of equally slow, but distinct, rate-limiting assembly pathways.
At low target numbers, fluctuations favor pathways that lead to excess nucleations, which irreversibly reduces the average final yield.
Even biochemically favorable assembly compartments, which can individually assemble structures at high yield, are susceptible to this stochastic yield reduction.
Hence, the spatial separation that benefits assembly at large target numbers can become detrimental when only few target structures are assembled.

Understanding that rate-limiting exchange dynamics are the origin of the stochastic yield catastrophe further provides us with strategies for mitigating it.
We show that high yield can be recovered by modifying the exchange dynamics such that larger structures exchange slower than smaller ones.
Importantly, reducing the exchange rates of larger structures does not affect the assembly efficiency at the mean-field level~\citep{angerpointnerDelayfacilitatedSelfassemblyCompartmentalized2025}, but drastically improves robustness against fluctuations at low target numbers.

We conclude that the stochastic behavior of an assembly model at low target numbers cannot be inferred from its mean-field limit.
Seemingly irrelevant details of the mean-field model, such as the structure size dependence of exchange rates, or adding an activation step to subunits~\citep{gartnerStochasticYieldCatastrophes2020}, can create or remedy stochastic yield catastrophes.
A common feature of such yield catastrophes at low target numbers is the emergence of competing rate-limiting assembly pathways whose random order impacts the final yield sensitively.

\section{Model and mean-field behavior}
\label{sec:model}

\begin{figure*}[ht]
    \centering
    \includegraphics[]{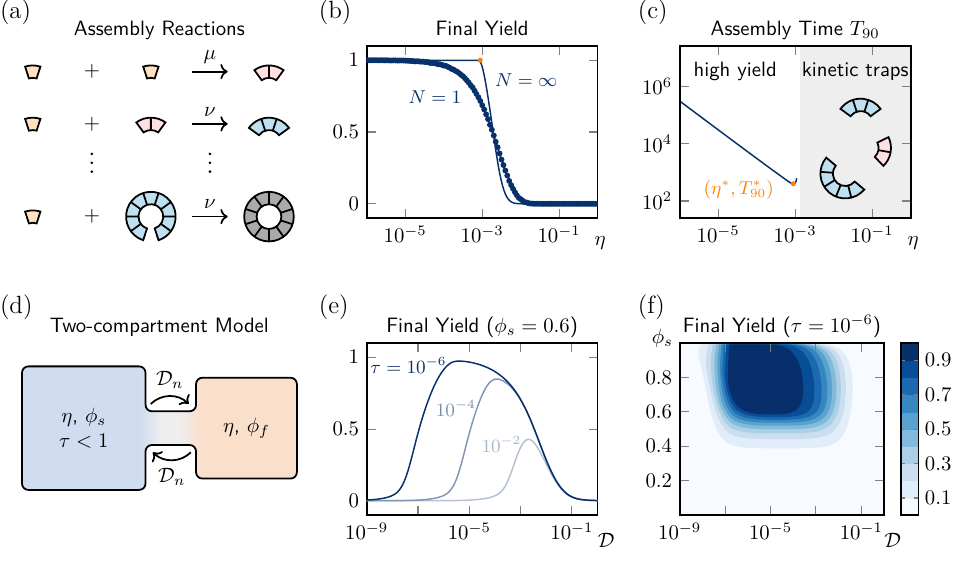}
    \caption{Recap of well-mixed and delay-facilitated self-assembly from Ref.\@~\citep{angerpointnerDelayfacilitatedSelfassemblyCompartmentalized2025} at structure size~${S=30}$. 
    (a)~Sketch of assembly reactions: nucleation at rate~$\mu$ and structure growth at rate~$\nu$.
    (b)~Final Yield in a well-mixed compartment as a function of the nucleation-to-growth ratio~$\eta$. 
    Low~$\eta$ is required for high yield; high~$\eta$ leads to kinetic traps.
    Bullets (${N=1}$) show stochastic simulations with~$S$ subunits, sufficient to assemble at most a single target structure.
    (c)~Corresponding assembly times~$T_{90}$. 
    Low~$\eta$ slows down the assembly; assembly is fastest at the optimal value of~$\eta^*$ (orange bullet).
    (d)~Sketch of the two-compartment model with exchange parameters~$\mathcal{D}_n$, relative compartment volumes~$\phi_\alpha$, and relative compartment reactivity~$\tau$.
    (e)~Final yield as a function of the structure-size-independent exchange parameter~${\mathcal{D}_n=\mathcal{D}}$ at varying relative compartment reactivities~$\tau$, with 
    fixed volume ratio $\phi_s=0.6$ and nucleation-to-growth ratio~$\eta=10\eta^*$.
    A parameter regime of improved yield for intermediate~$\mathcal{D}$ emerges.
    (f)~Corresponding contour plot for varying relative compartment volume~$\phi_s$ at fixed~${\tau=10^{-6}}$. 
    Individual panels reproduced and adapted from Figs.\@~1 and~2 in Ref.~\citep{angerpointnerDelayfacilitatedSelfassemblyCompartmentalized2025}.
    }
    \label{fig:recap_delay_facilitated_assembly}
\end{figure*}

To investigate stochastic effects at low target numbers in spatially organized self-assembly, we build on a minimal model that has been studied both in well-mixed environments~\citep{gartnerStochasticYieldCatastrophes2020} and in spatially organized two-compartment systems~\citep{angerpointnerDelayfacilitatedSelfassemblyCompartmentalized2025}.
In this section we introduce this minimal model and briefly recap the most important mean-field results from these studies; for a more detailed analysis we refer the reader to Refs.\@~\citep{gartnerStochasticYieldCatastrophes2020, angerpointnerDelayfacilitatedSelfassemblyCompartmentalized2025}.
We later compare these mean-field results to their stochastic counterparts at low target numbers (Sec.\@~\ref{sec:results}).

\subsection{Well-mixed self-assembly model}

As a baseline against which we can assess the effects of spatial organization and stochasticity, we consider a conceptual assembly model [Fig.\@~\ref{fig:recap_delay_facilitated_assembly}(a)] with two distinct pairwise interactions between identical subunits (e.g., individual proteins): \textit{nucleation} with rate~$\mu$ and \textit{growth} with rate~$\nu$.
Nucleation produces the smallest thermodynamically stable intermediate (here assumed to be dimers), which grows by sequential subunit addition until completing a one-dimensional ring of~$S$ subunits, a  minimal representation of closed structures such as viral capsid shells.
We assume all structures to be thermodynamically stable on the timescale of the assembly process (irreversible bindings) and that growth rates are independent of structure size.

The process starts from a well-mixed solution of free subunits at density~$\rho_0$, with no pre-formed structures, and terminates once all subunits are incorporated into structures.
Its performance is quantified in terms of the final yield~$Y$ and the assembly time~$T_{90}$.
The final yield measures the average fraction of subunits integrated at the end of this process and~$T_{90}$ is the time it takes to reach an average yield of 90\% (measured in units of the growth rate times the initial subunit density~$\rho_0$).
As a consequence of the irreversible bindings, the system is susceptible to \textit{kinetic traps} if nucleation is too fast compared to growth.
Consequently, the \textit{nucleation-to-growth ratio}~${\eta=\mu/\nu}$ has to be small to achieve high yields~[Fig.\@~\ref{fig:recap_delay_facilitated_assembly}(b) and Ref.\@~\citep{gartnerStochasticYieldCatastrophes2020}]---an instance of the well-known slow-nucleation principle~\citep{zlotnickTheoreticalModelSuccessfully1999, morozov_assembly_2009, hagan_understanding_2010, haganMechanismsKineticTrapping2011, keThreeDimensionalStructuresSelfAssembled2012, weiComplexShapesSelfassembled2012, reinhardtNumericalEvidenceNucleated2014, gartnerStochasticYieldCatastrophes2020}.
However, if nucleation is too slow ($\eta$ too small), the self-assembly is rate-limited by the slow nucleation and the assembly time increases~[Fig.\@~\ref{fig:recap_delay_facilitated_assembly}(c)].
Thus, there is an \textit{optimal} nucleation-to-growth ratio~$\eta^*$, where the assembly process is both effective (high yield) and efficient (low assembly time).

\subsection{Extension to two compartments}

The well-mixed results highlight the central role of the nucleation-to-growth ratio~$\eta$ for self-assembly performance.
In experimental and biological settings, however, nucleation and growth rates are set by microscopic particle properties such as interaction strengths or particle shapes, and are difficult to tune independently.
Therefore, the nucleation-to-growth ratio might be fixed to an undesirably large value~${\eta>\eta^*}$, requiring alternate strategies to achieve high-yield assembly.

One way to recover high-yield assembly without fine-tuning biochemical properties is through the spatial separation of reaction environments~\citep{haganSelfassemblyCoupledLiquidliquid2023, laha_chemical_2024, bartolucci_interplay_2024, frechetteComputerSimulationsShow2025, angerpointnerDelayfacilitatedSelfassemblyCompartmentalized2025}.
Here, we analyze a conceptual two-compartment model, that couples a high-reactivity compartment and a low-reactivity compartment via particle exchange~[Fig.\@~\ref{fig:recap_delay_facilitated_assembly}(d) and Ref.\@~\citep{angerpointnerDelayfacilitatedSelfassemblyCompartmentalized2025}].
Both compartments, which we also call \textit{fast} and \textit{slow}, respectively, share the same nucleation-to-growth ratio~$\eta$, but all reaction rates in the slow compartment are reduced by a constant factor~${\tau < 1}$.
Despite its simplicity, this two-compartment model captures the essential features of more complex spatially organized assembly scenarios~\citep{angerpointnerDelayfacilitatedSelfassemblyCompartmentalized2025}.

The state of each compartment~$\alpha \in \{f, s\}$ (for fast and slow, respectively) is described by the densities~$\sigma_{n,\alpha}$ of structures of size~$n$, measured in units of the initial subunit density~$\rho_0$. 
These densities evolve through three processes:
nucleation, growth and particle exchange.
Within each compartment~$\alpha$ the subunit density~($\sigma_{1,\alpha}$) is depleted by nucleation and growth; dimers~($\sigma_{2, \alpha}$) are created through nucleation; and larger structures of size~$n$ ($\sigma_{n, \alpha}$) are created through subunit attachment to smaller structures~($\sigma_{n-1, \alpha}$), until target structures~($\sigma_{S, \alpha}$) are completed.
Particle exchange relaxes density differences between compartment~$\alpha$ and~$\beta$ (with~${\alpha\neq\beta}$). 
In the deterministic mean-field limit, this yields~\citep{angerpointnerDelayfacilitatedSelfassemblyCompartmentalized2025}:
\begin{subequations}
\label{eq:model_ode}
\begin{alignat}{8}
        \partial_t \sigma_{1, \alpha} 
        &= \mathcal{D}_{1,\alpha}
        - \tau_\alpha \left[2\eta (\sigma_{1, \alpha})^2 
        +\sigma_{1, \alpha}\sum_{n=2}^{S-1}\sigma_{n,\alpha} \right] \,, 
        \label{eq:model_ode_1}
        \\[2mm]
        \partial_t \sigma_{2, \alpha} 
        &= \mathcal{D}_{2,\alpha} 
        + \tau_\alpha  \left[ \eta(\sigma_{1, \alpha})^2 {-} \sigma_{1, \alpha}\sigma_{2,\alpha} \right] \, ,
        \label{eq:model_ode_2}
        \\[2mm]
        \partial_t \sigma_{n, \alpha} 
        &= \mathcal{D}_{n,\alpha} 
        + \tau_\alpha \left[ \sigma_{n{-}1,\alpha} 
        - \sigma_{n,\alpha} \right] \sigma_{1, \alpha} \, ,
        \label{eq:model_ode_3}
        \\[2mm]
        \partial_t \sigma_{S, \alpha} 
        &= \mathcal{D}_{S,\alpha} 
        + \tau_\alpha \sigma_{1, \alpha} \sigma_{S-1,\alpha} \, ,
        \label{eq:model_ode_4}
\end{alignat}
with the relative reactivity~$\tau_\alpha$ in each compartment (we choose~${\tau_f=1}$ and ${\tau\equiv\tau_s < 1}$), and the inter-compartment exchange flux,
\begin{equation}
     \mathcal{D}_{n,\alpha} = \frac{\mathcal{D}_n (\sigma_{n,\beta} - \sigma_{n,\alpha})}{\phi_\alpha} \, ,
\end{equation}
\end{subequations}
from compartment~$\alpha$ to~$\beta$, depending on the structure size~$1\leq n \leq S$.

This assembly model features two classes of parameters.
The \textit{biochemical parameters}, i.e., the nucleation-to-growth ratio~$\eta$, the relative compartment reactivity~$\tau$ and the structure size~$S$ govern the assembly dynamics within each compartment.
The \textit{geometric parameters}~${\phi_\alpha=V_\alpha/V}$ and~$\mathcal{D}_n$, on the other hand, control the coupling between compartments. 
The relative compartment size~$\phi_\alpha$ measures the compartment volumes~$V_\alpha$ with respect to the total volume~${V=V_f+V_s}$ and the non-dimensional exchange parameter~$\mathcal{D}_n$ sets the exchange rate for structures of size~$n$ in units of the total volume and the reactive timescale of the fast compartment. 
Exchange is assumed to be symmetric between compartments, since asymmetric exchange merely rescales the effective compartment volumes~\citep{angerpointnerDelayfacilitatedSelfassemblyCompartmentalized2025}.
Henceforth, we set~$S=30$ as the target size does not qualitatively affect our findings (see Appendix~\ref{app:diff-target-sizes}) and use the geometric parameters~$\phi_\alpha$ and~$\mathcal{D}_n$ as the primary control parameters.

\subsection{Delay-facilitated self-assembly}

The dynamics described by Eq.~\eqref{eq:model_ode} have two clear limiting cases for size-independent exchange: 
vanishing exchange~${\mathcal{D}_n \rightarrow 0}$ decouples the compartments, while fast exchange~${\mathcal{D}_n \rightarrow \infty}$ effectively merges them into a single well-mixed compartment.
For large nucleation-to-growth ratios (e.g., ${\eta = 10\eta^*}$), both limiting cases are characterized by low final yields, as each individual compartment suffers from kinetic traps [Fig.~\ref{fig:recap_delay_facilitated_assembly}(e)]. 
In the intermediate exchange regime, however, numeric solutions of~Eq.\@~\eqref{eq:model_ode} reveal that high yield is recovered over an extended range of exchange parameters, provided the slow compartment is sufficiently slow~(${\tau\ll 1}$) and large enough~[${\phi_s > 0.5}$ in Figs.~\ref{fig:recap_delay_facilitated_assembly}(e)\nobreakdash--(f); Ref.\@~\citep{angerpointnerDelayfacilitatedSelfassemblyCompartmentalized2025}]. 
This yield recovery originates from a separation of timescales: If the reactivity in both compartments is sufficiently different~(${\tau \ll 1}$) the exchange dynamics can be chosen such that they are \textit{slow} compared to the assembly dynamics in the fast compartment, but \textit{fast} compared to the assembly in the slow compartment.
This separation of timescales ensures an effectively reduced nucleation-to-growth ratio, allowing for high final yields (this mechanism is explained in more detail in Sec.\@~\ref{sec:results}).
Whether and how the exchange parameters~$\mathcal{D}_n$ depend on the structure size affects these mean-field results only minimally (Appendix~\ref{app:size-dependent_exchange_mean_field}).
This recovery of high yield for intermediate exchange rates we called \textit{delay-facilitated self-assembly}~\citep{angerpointnerDelayfacilitatedSelfassemblyCompartmentalized2025}.

While delay-facilitated self-assembly offers a promising way to control assembly via geometric parameters rather than the less accessible intrinsic interaction strengths, it remains unclear whether it can sustain high yield in the presence of stochastic fluctuations at low target numbers.
In the following sections, we therefore analyze self-assembly in the two-compartment model \textit{beyond} the mean-field limit~[Eq.\@~\eqref{eq:model_ode}].

\section{Stochastic results}
\label{sec:results}

\begin{figure*}[ht]
    \includegraphics[]{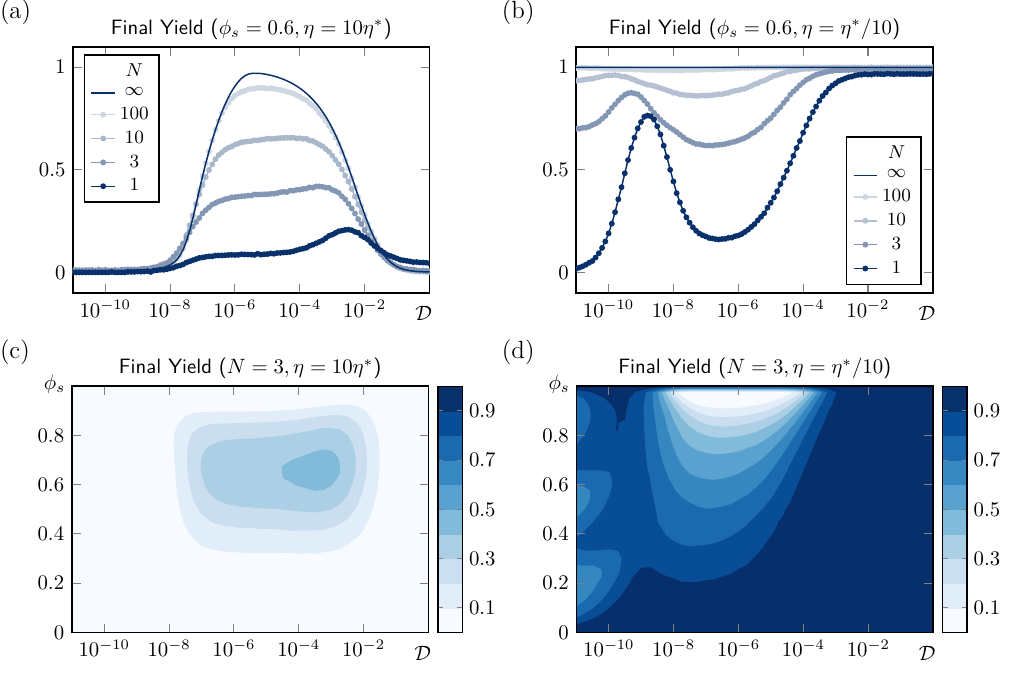}
    \caption{Low target numbers cause a stochastic yield catastrophe.
        (a)~Final yield for two coupled low-yield compartments (${\eta=10\eta^*}$) as a function of the structure-size-independent exchange parameter $\mathcal{D}$ for varying target numbers~$N$ at fixed initial density $\rho_0$, relative compartment volume~$\phi_s=0.6$ and relative reactivity~$\tau=10^{-6}$. 
        Solid lines (${N=\infty}$) correspond to numeric solutions of the mean-field equations~[Eq.\@~\eqref{eq:model_ode}]. 
        Reducing the target number~$N$ reduces the yield at intermediate exchange parameters~$\mathcal{D}$; at~${N=1}$, the yield almost vanishes, the signature of a stochastic yield catastrophe.
        (c)~Corresponding contour plot at fixed target number~${N=3}$ and varying relative compartment volume~$\phi_s$.
        (b),(d)~Same as (a) and~(c), but for two coupled high-yield compartments ($\eta=\eta^*/10$).
        The stochastic yield catastrophe persists even when each compartment in isolation supports high-yield assembly.
    }
    \label{fig:results_yield_catstrophe_2comp}
\end{figure*}

Previous work has shown that stochastic effects play only a minor role in the well-mixed scenario at low target numbers~\citep{gartnerStochasticYieldCatastrophes2020}.
In particular, even a single target structure can be assembled robustly if 
the nucleation-to-growth ratio~$\eta$ is small enough~[Fig.\@~\ref{fig:recap_delay_facilitated_assembly}(b)].
However, due to its spatial separation, the two-compartment model features different sources of stochasticity:
the initial distribution of subunits, the assembly dynamics within each individual compartment, and the exchange dynamics between compartments.

\subsection{Stochastic yield catastrophe}

To investigate how stochastic fluctuations affect the final yield of the two-compartment assembly model [Figs.\@~\ref{fig:recap_delay_facilitated_assembly}(a),(d)]
at low target numbers~$N$, we employ the \textit{stochastic simulation algorithm}~\citep{Gillespie1976, Gillespie.2007}.
This allows us to directly simulate the discrete assembly process, rather than solving the deterministic mean-field equations.
We generate multiple realizations of the underlying stochastic master equation with a total number of~${N_0 =SN}$ subunits, the stoichiometric amount required to form exactly~$N$ target structures.
The random initial distribution of subunits among compartments and the per-capita reaction rates are chosen to match the particle densities and intrinsic rate constants of Eq.\@~\eqref{eq:model_ode}, respectively, ensuring that the stochastic system converges to the same physical model in the deterministic limit~${N \to \infty}$ (see Appendix~\ref{app:master_equation} for the definition of the stochastic process, the number of realizations used per simulation, and error estimations).

We first focus on the simplest case of structure size-independent exchange parameters~$\mathcal{D}_n = \mathcal{D}$, since the size dependence of the exchange dynamics has only a weak effect on the mean-field assembly dynamics~(see Appendix~\ref{app:size-dependent_exchange_mean_field} and Ref.\@~\citep{angerpointnerDelayfacilitatedSelfassemblyCompartmentalized2025}).
Using exemplary values of~$\tau=10^{-6}$ and~$\phi_s = 0.6$ for the relative compartment reactivities and volumes, we compare stochastic simulations at different target numbers~$N$ to the deterministic limit [Fig.\@~\ref{fig:results_yield_catstrophe_2comp}(a)].
We find that for large target numbers (${N=100}$), delay-facilitated assembly performs well at intermediate exchange parameters (${10^{-7} < \mathcal{D} < 10^{-3}}$), in close agreement with the deterministic mean-field result.
In contrast, the efficacy of delay-facilitated self-assembly, and thereby the maximally achievable yield, is strongly reduced for lower target numbers: 
While for~${N=10}$ a yield of approximately 60\% is still possible, it almost vanishes if only enough subunits for a single target structure are present~[$N=1$; Fig.\@~\ref{fig:results_yield_catstrophe_2comp}(a)].
This yield reduction also persists at different relative compartment volumes.
Comparing Figs.\@~\ref{fig:recap_delay_facilitated_assembly}(f)~(${N=\infty}$) and~\ref{fig:results_yield_catstrophe_2comp}(c)~($N=3$), one observes that both for high and for low target numbers, intermediate values of the exchange parameter~$\mathcal{D}$ lead to increased yield for different relative compartment volumes; however, the maximal yield is much lower at low target numbers.
These results are in stark contrast to the well-mixed assembly process, where the final yield changes only weakly with the total number of subunits~[Fig.\@~\ref{fig:recap_delay_facilitated_assembly}(b)].

The reduction in final yield at low target numbers and intermediate exchange parameters is a stochastic effect.
It even persists for two coupled high-yield compartments with~${\eta = \eta^*/10}$.
For such compartments, a final yield of almost~100\% at high target numbers is reached, regardless of the exchange parameter~$\mathcal{D}$~[Fig.\@~\ref{fig:results_yield_catstrophe_2comp}(b)].
Since the intrinsic assembly dynamics in each compartment are efficient and not susceptible to stochastic fluctuations, one might expect that perfect yield can be reached at low target numbers as well.
However, at low target numbers, even for two coupled high-yield compartments, two parameter regimes of reduced yield emerge~[Figs.\@~\ref{fig:results_yield_catstrophe_2comp}(b),(d)]:
For effectively isolated compartments~($\mathcal{D}\rightarrow0$) the final yield is coupled to the commensurability between target numbers and the compartment volumes~(Appendix~\ref{app:decoupled_compartments_yield}).
This is not an inherently stochastic effect and therefore not the focus of this study.
Much more importantly, in the regime where delay-facilitated assembly is expected to promote self-assembly (intermediate exchange parameters), the final yield is strongly reduced at low target numbers~[Figs.\@~\ref{fig:results_yield_catstrophe_2comp}(b),(d)].

The observed \textit{stochastic yield catastrophe} therefore occurs independently of the nucleation-to-growth ratio in each compartment.
Instead, it originates from the spatial separation of reaction environments, which introduces additional noise sources that  interact with the assembly dynamics at low particle numbers.
In the following section, we analyze the dynamic origin of this effect in detail.

\begin{figure}[t!]
    \includegraphics[]{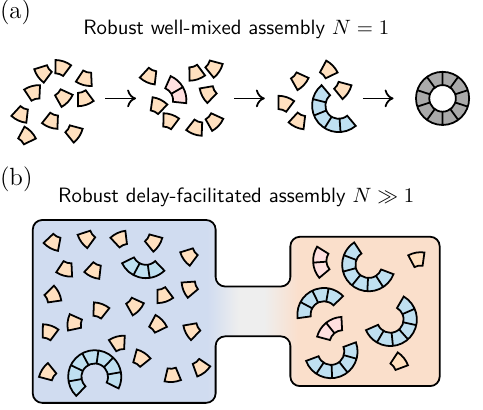}
    \caption{Mechanisms for robust assembly.
    (a)~Snapshots during well-mixed assembly~${N=1}$ and low nucleation-to-growth ratio~${\eta<\eta^*}$.
    The slow nucleation rate fixes the order of nucleation and growth reactions, leading to high final yield.
    (b)~Typical assembly state during delay-facilitated self-assembly in the two-compartment system at intermediate exchange parameter~$\mathcal{D}$ and high target numbers.
    The fast compartment (right) contains many growing structures and few free subunits; this suppresses nucleation and facilitates high final yield.
    }
    \label{fig:avoid_catastrophe}
\end{figure}

\subsection{Dynamic origin of the delay-induced stochastic yield catastrophe}
\label{sec:origin_of_catastrophe}

\begin{figure*}[ht]
    \includegraphics[]{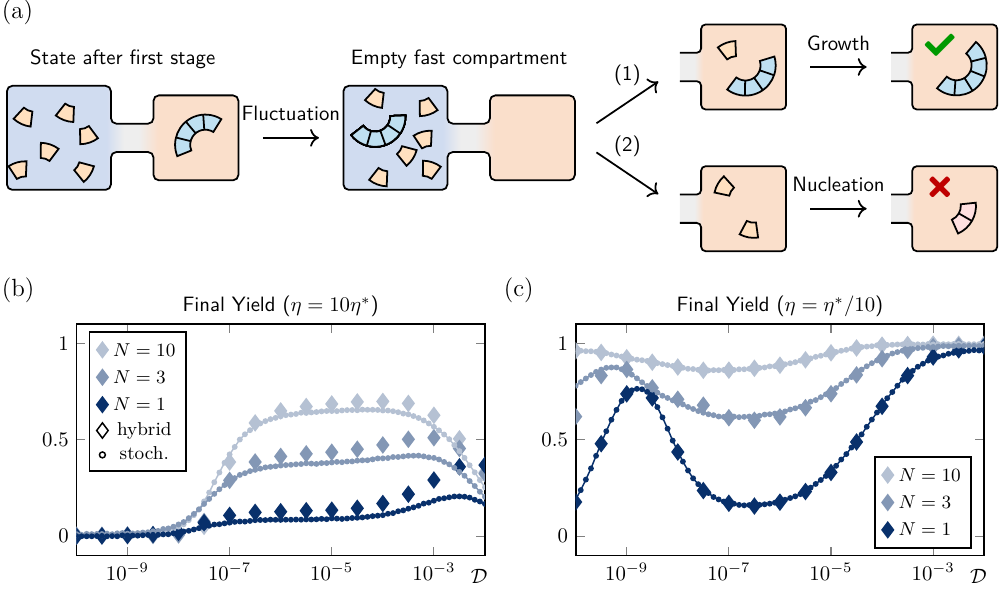}
    \caption{
    Mechanism leading to the delay-induced stochastic yield catastrophe in the two-compartment system.
    (a) Sketch of the mechanism (depicted for~${N=1}$). 
    Left: Typical state after the first assembly stage. 
    Middle: In the second stage, fluctuations induced by particle exchange can leave the fast compartment without growing structures. 
    Right: Depending on the order of subsequent particle exchanges, the next reaction can either lead to productive structure growth~[pathway~(1)] or excess nucleation [pathway~(2)].
    (b),(c)~Final yield obtained from fully stochastic simulations (small bullets) and from deterministic-stochastic hybrid simulations (first stage deterministic, second stage stochastic; large diamond markers) for different target numbers~$N$ as a function of the structure-size-independent exchange parameter~$\mathcal{D}$, in two coupled low-yield compartments [${\eta=10\eta^*}$;~(b)] and two coupled high-yield compartments [${\eta=\eta^*/10}$;~(c)] .
    The close agreement between fully stochastic and hybrid simulations shows that the second assembly-stage is responsible for the delay-induced stochastic yield catastrophe.
    }
    \label{fig:mechanism}
\end{figure*}

To understand why the two-compartment model exhibits a stochastic yield catastrophe at low target numbers while the well-mixed model does not, and which source of stochasticity is responsible, we analyze the assembly dynamics in terms of the number of qualitatively different types of rate-limiting events.
For the well-mixed assembly at low nucleation-to-growth ratios~$\eta \ll \eta^*$, nucleation is the only type of rate-limiting event: 
every nucleation is therefore followed by a cascade of fast growth steps, fixing the order of reactions through a separation of timescales and leading to high final yields---even at low target numbers~[Fig.\@~\ref{fig:avoid_catastrophe}(a) and Ref.\@~\citep{gartnerStochasticYieldCatastrophes2020}].

In contrast, delay-facilitated self-assembly operates in the regime of a two-fold timescale separation where inter-compartmental exchange is much faster than the assembly dynamics in the slow compartment, but much slower than the assembly in the fast compartment.
This two-fold separation gives rise to two distinct assembly stages~\citep{angerpointnerDelayfacilitatedSelfassemblyCompartmentalized2025}.
During the first stage, the assembly dynamics are dominated by the assembly reactions in the fast compartment, while only a few particles exchange between compartments or react in the slow compartment.
As a result, the system approaches a state in which the fast compartment contains relatively many unfinished structures but few free subunits, whereas the slow compartment contains mostly free subunits and few, if any, growing structures~[Fig.\@~\ref{fig:avoid_catastrophe}(b)].

In the second stage, the assembly dynamics change qualitatively:
Due to a lack of free subunits, assembly in the fast compartment is rate-limited by the slow influx of subunits. 
As long as reactions in the slow compartment are slow enough~(${\tau\ll 1}$), this exchange-limited assembly in the fast compartment is dominant.
The combination of fast assembly and slow subunit influx keeps the \textit{average} subunit-to-structure ratio in the fast compartment low, thereby suppressing nucleation relative to growth.
Thus, at least at high target numbers, nucleation events remain rare and high yields can be achieved.
In contrast to the well-mixed scenario, however, this exchange-limited assembly process with a low effective nucleation-to-growth ratio involves two types of comparably slow events when subunits and structures exchange at comparable rates: the exchange of subunits \textit{and} the exchange of structures.
These events occur in a random order, causing fluctuations in the subunit-to-structure ratio in the fast compartment and thereby in the balance between nucleation and growth.

To understand how a second type of slow event induces a stochastic yield catastrophe, it is instructive to analyze both assembly stages for two coupled high-yield compartments~(${\eta<\eta^*}$) with~${N=1}$, i.e., with only enough subunits for a single target structure.
As discussed above, the separation of timescales between particle exchange and the assembly reactions means that the first assembly stage is well approximated by the assembly in an isolated well-mixed fast compartment, the outcome of which is robust against stochastic fluctuations.
After the first stage, the fast compartment therefore typically contains a single growing structure, while the slow compartment contains only free subunits, largely independently of the initial random partitioning of subunits between compartments [left panel in Fig.\@~\ref{fig:mechanism}(a)].

In the following second assembly stage, the order in which particles exchange between compartments determines whether the next reaction will be a nucleation or growth event.
In particular, the only growing structure may, and at some point most likely will, exchange to the slow compartment, leaving the fast compartment without any growing structure [center panel in Fig.\@~\ref{fig:mechanism}(a)].
In this state, two different pathways are possible:
Either the next particles to exchange to the fast compartment are the structure and a subunit, leading to structure growth, or two subunits exchange to the fast compartment in succession and nucleate~[pathways~(1) and (2) in Fig.\@~\ref{fig:mechanism}(a), respectively].
For~${N=1}$, this single excess nucleation is already enough to prevent the completion of the target structure.
Importantly, within the timescale-separated regime considered here, whether the next reaction is productive growth or detrimental nucleation is controlled primarily by the random order of the slow exchange events, rather than by the values of the exchange parameter~$\mathcal{D}$ or the microscopic nucleation-to-growth ratio~$\eta$.
At large target numbers, however, the random ordering of exchange events does not affect the assembly outcome in the same drastic way:
The average structure-to-subunit ratio in the fast compartment remains high and a sequence of exchange events that sufficiently depletes this ratio becomes increasingly unlikely with growing~$N$.

Although the first assembly stage is subject to stochastic fluctuations as well, these fluctuations cannot account for the stochastic yield catastrophe observed for two high-yield compartments [Figs.\@~\ref{fig:results_yield_catstrophe_2comp}(b) and~(d)], since the approximately well-mixed assembly in the fast compartment remains robust even at low particle numbers.
Instead, the second stage introduces a qualitatively new source of stochasticity through the random ordering of comparably slow exchange events.

To quantitatively compare the impact of these different sources of stochasticity on the final yield, we isolate the stochastic effects in the second stage using a hybrid simulation scheme:
The first assembly stage is approximated by mean-field simulations of an isolated fast compartment, and its final state serves as the initial condition for a stochastic simulation of the second stage.
The final state of the first stage is discretized for each~$N$ by rounding the mean-field particle densities in the fast compartment to their closest discrete analogue and placing the remaining material as free subunits in the slow compartment (see Appendix~\ref{app:two_stage_simulations}).
For sufficiently slow particle exchange, this hybrid scheme approximates the actual assembly process with suppressed stochastic fluctuations during the first stage, including the fluctuations in initial conditions.
The resulting final yields agree well with those from the fully stochastic simulations [compare the diamond and circular markers in Figs.\@~\ref{fig:mechanism}(b)\nobreakdash--(c)].
This shows that the stochasticity of the second assembly stage alone is sufficient to produce the stochastic yield catastrophe.
It is precisely the random order of two comparably slow exchange events---subunit exchange versus structure exchange---that determines whether the next reaction in the fast compartment is productive growth or excess nucleation.
The same timescale separation that enables delay-facilitated assembly at high target numbers thus also introduces an additional rate-limiting exchange pathway that reduces yield at low target numbers.
Because this stochastic yield catastrophe is caused by the delayed, exchange-limited supply of material between compartments, we refer to it as a \textit{delay-induced stochastic yield catastrophe}.

\subsection{Controlled particle exchange alleviates stochastic yield catastrophe}
\label{sec:control-exchange}

\begin{figure*}[ht]
    \includegraphics[]{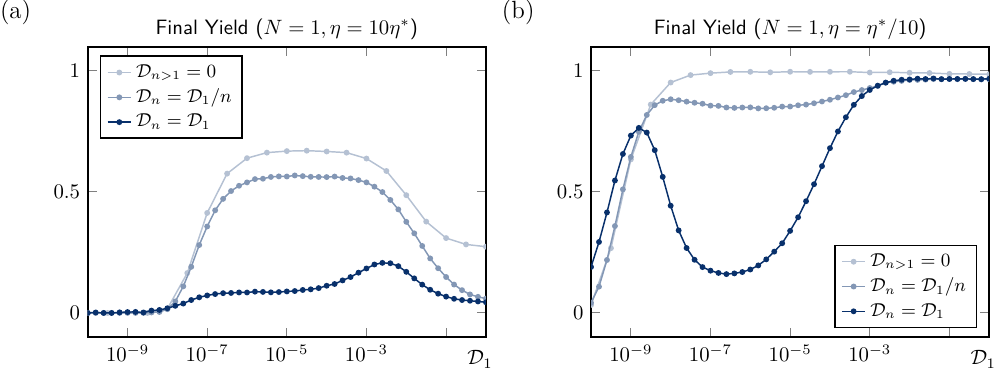}
    \caption{Different exchange dynamics can alleviate the stochastic yield catastrophe. 
    (a),(b)~Final yield for two coupled low-yield [${\eta=10\eta^*}$;~(a)] or high-yield [${\eta=\eta^*/10}$;~(b)] compartments at~${N=1}$, as a function of the subunit exchange parameter~$\mathcal{D}_1$, for different exchange dynamics: no structure exchange~($\mathcal{D}_{n>1}=0$), slower exchange for larger structures~(${\mathcal{D}_n=\mathcal{D}_1/n}$) and size-independent exchange~(${\mathcal{D}_n=\mathcal{D}_1}$). 
    Relative compartment volumes are fixed to~$\phi_s=0.6$ and the relative compartment reactivity to~${\tau=10^{-6}}$.
    The delay-induced stochastic yield catastrophe is alleviated when structure exchange is slower than subunit exchange.  
    }
    \label{fig:controlled_exchange}
\end{figure*}

In the previous section we identified the cause of the delay-induced stochastic yield catastrophe: the random ordering of comparably slow subunit and structure exchange events.
This understanding allows us to formulate strategies to alleviate the yield reduction at low target numbers.
One natural design principle is to break the equivalence between subunit and structure exchange in a way that promotes structure growth and suppresses excess nucleation.
Excess nucleation is driven by fluctuations that lead to below-average structure numbers in the fast compartment---e.g., a fast compartment without growing structures [Fig.\@~\ref{fig:mechanism}(a)].
To avoid such fluctuations, structure exchange should be slower than subunit exchange, i.e., ${\mathcal{D}_n<\mathcal{D}_1}$ in Eq.\@~\eqref{eq:model_ode}.
Two ways this can be achieved are through size-dependent exchange parameters~${\mathcal{D}_n=\mathcal{D}_1/n}$ akin to Stokes-Einstein diffusion coefficients, or by eliminating structure exchange entirely~(${\mathcal{D}_{n>1}=0}$).

Numerical simulations of two compartments with restricted exchange dynamics show that even for very low target numbers, most of the stochastic yield reduction is alleviated when structure exchange is suppressed entirely~(${\mathcal{D}_{n>1}=0}$), or when the exchange parameter is size-dependent~(${\mathcal{D}_n=\mathcal{D}_1/n}$; Fig.\@~\ref{fig:controlled_exchange}).
Intermediate exchange rates again facilitate a significantly improved yield for two low-yield compartments~[Fig.\@~\ref{fig:controlled_exchange}(a)]. 
Moreover, as long as reactions in the slow compartment can be neglected, two coupled high-yield compartments~[${\eta<\eta^*}$; Fig.\@~\ref{fig:controlled_exchange}(b)] recover the mean-field value of almost 100\% yield. 
The remaining deviation from the mean-field results for low-yield compartments with~${\eta=10\eta^*}$ can be traced to stochastic variations of the system's state after the first stage, consistent with the two-stage analysis of Sec.\@~\ref{sec:origin_of_catastrophe}.
For example, for~${N=1}$, a second nucleation in the fast compartment during the first stage prevents the assembly of the target structure; the maximal final yield is therefore capped by the probability of avoiding such an event, which is approximately 67\% (see Appendix~\ref{app:stochasticity_without_structure_exchange}). 

Importantly, for this controlled exchange scenario and in contrast to the delay-induced stochastic yield catastrophe~(Fig.\@~\ref{fig:results_yield_catstrophe_2comp}), delay-facilitated self-assembly remains functional, albeit at a slightly reduced efficacy.
In particular, stochastic fluctuations no longer cause a substantial yield reduction in two coupled high-yield compartments.
This follows from breaking the equivalence of previously equally slow exchange events.
Hence, the random ordering of exchange events is now biased toward subunit exchange, thereby favoring structure growth over excess nucleation.
These findings again highlight the difference between the mean-field assembly dynamics at high target numbers and the stochastic assembly dynamics at low target numbers:
While size-dependent exchange dynamics do not affect the mean-field dynamics of delay-facilitated self-assembly qualitatively, they strongly affect the severity of the delay-induced stochastic yield catastrophe.

\section{Discussion and Outlook}

\subsection{Summary}

In this work, we established two main results.
First, timescale separation between the assembly dynamics within spatial compartments and particle exchange is required for delay-facilitated assembly, but induces a stochastic yield catastrophe at low target numbers, even when each compartment in isolation supports robust assembly.
Fluctuations in the number of growing structures in the high-reactivity~(fast) compartment enhance the average rate of nucleating excess structures compared to the structure growth rate.
Whether an individual subunit entering the fast compartment contributes to excess nucleation or structure growth depends sensitively on the random order of previous slow exchange events.
In both high- and low-yield compartments, previous exchange of unfinished structures from the fast to the slow compartment depletes the fast compartment of growing structures and an arriving free subunit is more likely to nucleate an excess structure, thereby reducing the overall yield.

Second, this delay-induced stochastic yield catastrophe can be alleviated by controlling the exchange dynamics so that unfinished structures remain longer in the fast compartment.
Specifically, modifying the exchange dynamics, for instance by letting only subunits exchange between compartments or by reducing the exchange rates proportionally to the structure size, alleviates most of the observed detrimental stochastic effects.

\subsection{Generalizations to different assemblies and geometries}

\begin{figure*}[ht]
    \includegraphics[]{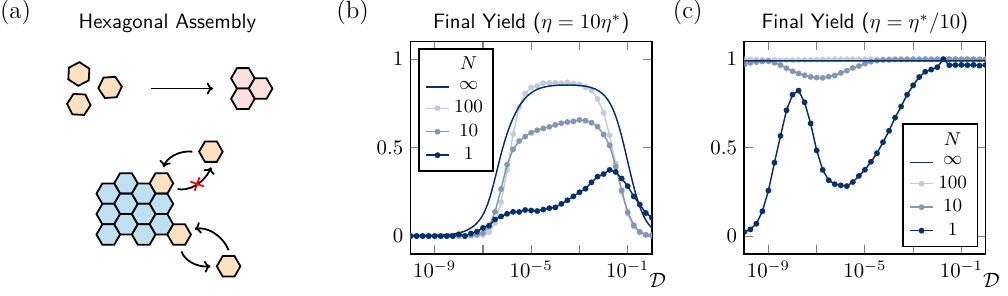}
    \caption{
    Assembly of two-dimensional subunits exhibit the same phenomenology. 
    (a)~Sketch of assembly reactions of hexagonal subunits.
    (b),(c)~Final yield for coupled low-yield compartments [${\eta=10\eta^*}$;~(b)] and high-yield compartments [${\eta=\eta^*/10}$;~(c)] at different target numbers~$N$ as a function of the structure-size-independent exchange parameter~$\mathcal{D}$. 
    Relative compartment volumes~$\phi_s=0.6$ and reactivities~$\tau=10^{-6}$ are fixed.
    Solid lines (${N=\infty}$) correspond to numerical solutions of effective deterministic mean-field equations (Appendix~\ref{app:hexagonal_assembly}). 
    Self-assembly of hexagonal-shaped subunits in two compartments also exhibits a delay-induced stochastic yield catastrophe, which therefore generalizes from one-dimensional to hexagonal subunits.
    }
    \label{fig:hexagons}
\end{figure*}

\begin{figure}[ht]
    \includegraphics[]{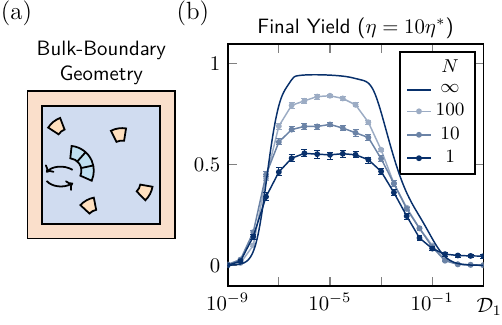}
    \caption{
    Diffusion-limited exchange recovers yield in a spatially extended system. 
    (a)~Sketch of a low-reactivity bulk coupled to a high-reactivity boundary.
    (b)~Final yield for a low-yield bulk and boundary (${\eta=10\eta^*}$) at different target numbers~$N$ as a function of the subunit bulk diffusion parameter~$\mathcal{D}_1$, with~$\mathcal{D}_n = \mathcal{D}_1/n$. 
    The relative compartment volume and reactivity are fixed to~$\phi_s=0.6$ and~$\tau=10^{-6}$. 
    The bulk size and boundary diffusion are kept constant; for more details about the spatially extended simulations and how all parameters are defined, see Appendix~\ref{app:cytosol_membrane_assembly}.
    Error bars indicate the $99\%$ confidence interval around each data point, see Appendix~\ref{app:sample_size_criterion}.
    }
    \label{fig:2d_membrane}
\end{figure}

In biological systems and synthetic applications, size-selective exchange can be implemented in different ways.
For example, strong size selection can occur through narrow pores in bacterial microcompartments~\citep{Chowdhury2015SelectiveProteinShells, Lee2017ShellDiffusionMutation, chowdhury_engineering_2019, tasneem_how_2022}, species-specific chaperone-induced transport~\citep{Phan.2018, Xing.2018}, or stronger binding affinities of larger structures to a reactive surface~\citep{khmelinskaia_control_2018}.
However, as we have shown, even much milder modifications of structure exchange, e.g., assuming the exchange rates of structures decrease proportionally to their size, can help to keep excess nucleations rare.
Such size-dependent diffusion rates often apply to particles in solution due to the Stokes-Einstein relation, but also membrane binding affinity can similarly increase with the oligomerization state~\citep{Hu.2002, Lang.2026}.

Furthermore, biological and synthetic assembly systems usually feature more complex structures or spatial organization than the linear assembly in our conceptual two-compartment model.
As long as the particular system of interest is in line with the basic assumptions of our model---e.g., coupled reaction environments with different reactivities---we predict that the same type of stochastic yield catastrophe emerges and that controlling particle exchange can alleviate it.

For instance, we can conceptually describe the assembly of two-dimensional structures such as virus capsids~\citep{Zlotnick2011VirusCapsids, Perlmutter2015VirusCapsids} or DNA bricks~\citep{siglProgrammableIcosahedralShell2021, monferrerDNAOrigamiTraps2023} by an effective model of hexagonal subunits featuring binding as well as detachment reactions [Fig.\@~\ref{fig:hexagons}(a)].
Despite its additional microscopic complexity, the mean-field dynamics of the hexagonal assembly model can be mapped to effective nucleation and growth reactions~\citep{Gartner2024}.
Importantly, we also observe the same stochastic effects at low target numbers in spatially organized hexagonal assembly [Fig.\@~\ref{fig:hexagons}(b),(c)], since the same rate-limiting assembly pathways emerge and can be influenced by modifying structure exchange.
Similarly, assembly compartments in biological cells need not be well-mixed bulk volumes, but can instead be a low-reactivity bulk coupled to a high-reactivity membrane~\citep{rosierProximityinducedCaspase9Activation2020, Leonard2023MembranesChangeReactionKinetics}, where particle exchange is governed by membrane affinity and bulk diffusion [Fig.\@~\ref{fig:2d_membrane}(a)].
In an example simulation of such a coupled cytosol-membrane system, we observe that robustness against fluctuations at low target numbers emerges naturally from size-dependent bulk diffusion, in line with our predictions [Fig.\@~\ref{fig:2d_membrane}(b)].
For more details about the hexagonal assembly simulations and the cytosol-membrane system, see Appendix~\ref{app:hexagonal_assembly} and~\ref{app:cytosol_membrane_assembly}, respectively.

\subsection{Outlook}

Our findings show that self-assembly strategies relying on spatial separation, in particular delay-facilitated assembly, can lead to a stochastic yield catastrophe.
This susceptibility to stochastic effects is a direct result of the spatially separated assembly dynamics and cannot be inferred from the well-mixed assembly model, which is robust against stochastic fluctuations.
Moreover, we have shown that stochastic robustness is sensitive to the size dependence of exchange rates, even though this dependence has little qualitative effect on the deterministic mean-field dynamics.
These results highlight that inferring the effect of stochastic fluctuations on assembly at low target numbers from either a well-mixed or a mean-field analysis is not generally possible.
To predict whether or not the final yield is robust against fluctuations, it is therefore crucial to understand the rate-limiting microscopic assembly pathways.

From a broader perspective, our results suggest that stochastic yield catastrophes arise when an assembly strategy introduces additional slow pathways whose random ordering can redirect the system away from productive growth.
Such pathways are particularly consequential at low target numbers, where individual exchange, activation, or nucleation events can change the subsequent fate of the entire assembly process.
This observation is in line with previous work that studied the robustness of different assembly strategies for single- and multi-component assemblies~\citep{gartnerStochasticYieldCatastrophes2020}.
The added complexity in the assembly state space by having different subunits revealed that one strategy (``activation scenario'') exhibits a stochastic yield catastrophe, while another (``nucleation scenario'') does not.
The origin of the stochastic yield catastrophe in the activation scenario is similar to the situation discussed here, namely the competition of multiple slow reaction pathways occurring in a random order.

The broader challenge is therefore not complexity per se, but the combination of low occupation numbers with multiple competing rate-limiting pathways.
In a large state space, individual species or intermediates may be represented by only few particles, so that rare sequences of events can have disproportionate consequences for the final outcome.
If such sequences favor excess nucleation, misordered activation, or other non-productive pathways, the effective requirements for high-yield assembly can be violated even when the corresponding mean-field dynamics appear favorable.
Robust low-target-number assembly therefore requires control mechanisms that suppress detrimental slow pathways or bias their ordering toward productive growth.

Overall, our work underlines that robust assembly at low target numbers cannot be assessed from deterministic mean-field behavior alone.
Moreover, even assembly models that are intrinsically robust in a well-mixed environment can become fluctuation-sensitive once coupled to additional spatial or internal degrees of freedom.
A common fingerprint of stochastic yield catastrophes is the presence of multiple competing rate-limiting pathways whose random ordering determines whether the system proceeds through productive growth or excess nucleation.
Identifying and controlling these pathways provides a practical route toward designing assembly strategies that remain efficient not only at large target numbers, but also in the low-target-number regimes relevant for many biological and synthetic systems.

\begin{acknowledgments}
We thank Florian Raßhofer and Jan Willeke for their critical review of the manuscript.
This work was supported by the Deutsche Forschungsgemeinschaft (DFG, German Research Foundation) under Germany’s Excellence Strategy through the Excellence Clusters ORIGINS (EXC-2094–390783311) and BioSysteM (EXC3092/1-533751719), and by the European Research Council (ERC) under the European Union’s Horizon Europe programme (Cell-Geom, Grant Agreement No. 101097810).
\end{acknowledgments}

\appendix

\section{Numerical simulations}
\label{app:numerics}

All numerical simulations used for this work were implemented in the Julia programming language.
Differential equations for the mean-field results in Figs.~\ref{fig:recap_delay_facilitated_assembly}, \ref{fig:results_yield_catstrophe_2comp}, \ref{fig:mechanism}, and \ref{fig:hexagons}--\ref{fig:appendix:effect_doesnt_depend_on_structure_size} were performed with the package DifferentialEquations.jl~\citep{rackauckas2017differentialequations} using an implicit Runge-Kutta method (ESDIRK).
The remaining results are obtained using our own implementation of the stochastic simulation algorithm (see Appendices~\ref{app:master_equation}, \ref{app:hexagonal_assembly}, and \ref{app:cytosol_membrane_assembly}).
For the bootstrap estimates of measurement uncertainties (see Appendix~\ref{app:sample_size_criterion}) we use the package Bootstrap.jl~\citep{juliangehringJuliangehringBootstrapjlBootstrap2023}.
All code used in this work is available on Zenodo~\citep{zenodo}.

\section{Chemical master equation and its parameters}
\label{app:master_equation}
The mean-field equations Eq.~\eqref{eq:model_ode} describe the evolution of the expectation values of an underlying stochastic process whose realizations are sampled in our simulations using Gillespie's stochastic simulation algorithm~\citep{Gillespie1976, Gillespie.2007}. 
In this appendix we state the chemical master equation that defines this stochastic process and relate its microscopic rate constants to the parameters of the mean-field model.

\subsection{State space and elementary reactions}

The system is characterized at time $t$ by the integer-valued vector of
structure counts,
\begin{equation}
\mathbf{N}(t) = \bigl\{ N_{n,\alpha}(t) \bigr\}_{\alpha\in\{f,s\},\;n=1,\dots,S} \, ,
\label{eq:appA_state}
\end{equation}
where $N_{n,\alpha}$ is the number of structures of size $n$ in
compartment $\alpha$. Compartment $\alpha$ has volume
$V_\alpha = \phi_\alpha V$; the total volume $V = S N / \rho_0$ is fixed
by the requirement that the system contain exactly enough material to
form $N$ target structures at the prescribed initial monomer
density~$\rho_0$.

For each compartment $\alpha\in\{f,s\}$, we distinguish three different classes of reactions:
\begin{subequations}
\label{eq:appA_reactions}
\begin{align}
    &\text{(i)\ nucleation:}\quad
    2\,X_{1,\alpha}\;\xrightarrow{\,c^{\mathrm{nuc}}_{\alpha}\,}\;X_{2,\alpha} \,,
    \\
    &\text{(ii)\ growth (}2\!\leq\!n\!<\!S\text{):} 
    \nonumber\\
    &\qquad
    X_{1,\alpha}+X_{n,\alpha}\;\xrightarrow{\,c^{\mathrm{grow}}_{\alpha}\,}\;X_{n+1,\alpha} \,,
    \\
    &\text{(iii)\ exchange (}\beta\!\neq\!\alpha\text{):}\quad
   X_{n,\alpha}\;\xrightarrow{\,c^{\mathrm{exch}}_{n,\alpha}\,}\;X_{n,\beta} \, .
\end{align}
\end{subequations}
The associated Gillespie propensities~$a_r(\mathbf{N})$---the
probabilities per unit time that a reaction of type~$r$ occurs when the system is
in state $\mathbf{N}$---follow as,
\begin{subequations}
\label{eq:appA_propensities}
\begin{align}
    a^{\mathrm{nuc}}_{\alpha}(\mathbf{N}) &= c^{\mathrm{nuc}}_{\alpha}\,\frac{N_{1,\alpha}\,(N_{1,\alpha}-1)}{2},\\[2pt]
    a^{\mathrm{grow}}_{n,\alpha}(\mathbf{N}) &= c^{\mathrm{grow}}_{\alpha}\, N_{1,\alpha}\, N_{n,\alpha},\\[2pt]
    a^{\mathrm{exch}}_{n,\alpha}(\mathbf{N}) &= c^{\mathrm{exch}}_{n,\alpha}\, N_{n,\alpha}.
\end{align}
\end{subequations}
Each reaction $r$ is further characterized by a stoichiometric change vector $\boldsymbol{\Delta}_r$ that updates the state as
$\mathbf{N}\!\to\!\mathbf{N}+\boldsymbol{\Delta}_r$; 
for example $\boldsymbol{\Delta}_{(\mathrm{nuc},\alpha)}=-2\,\mathbf{e}_{1,\alpha}+\mathbf{e}_{2,\alpha}$, where $\mathbf{e}_{n,\alpha}$ is the unit vector along the $(n,\alpha)$ component of $\mathbf{N}$.

\subsection{Chemical master equation}

Let $P(\mathbf{N},t)$ denote the probability that the system occupies state $\mathbf{N}$ at time $t$. 
From the state change propensities~Eq.\@\eqref{eq:appA_propensities} and the stoichiometric change vectors, the chemical master equation follows as,
\begin{align}
    \partial_t P(\mathbf{N},t) &= \sum_r \Big[\, a_r(\mathbf{N}-\boldsymbol{\Delta}_r)\,P(\mathbf{N}-\boldsymbol{\Delta}_r,t) \nonumber\\
     &\qquad\quad -\, a_r(\mathbf{N})\,P(\mathbf{N},t)\,\Big],
    \label{eq:appA_master}
\end{align}
where the sum runs over all reactions ${r \in \{(\mathrm{nuc},\alpha),(\mathrm{grow},n,\alpha),(\mathrm{exch},n,\alpha)\}}$ with $\alpha \in \{f,s\}$ and $n$ in the admissible range. 
The Gillespie algorithm then samples stochastically exact trajectories of~Eq.\@~\eqref{eq:appA_master} by drawing, in each step, the next reaction time from an exponential distribution with rate~$a_0(\mathbf{N})=\sum_r a_r(\mathbf{N})$ and the next reaction type, each with probability~$a_r(\mathbf{N})/a_0(\mathbf{N})$.

\subsection{Mapping to the mean-field parameters}

The microscopic rate constants $c^{\mathrm{nuc}}_{\alpha}$, $c^{\mathrm{grow}}_{\alpha}$, and $c^{\mathrm{exch}}_{n,\alpha}$ are fixed by requiring that the mean of the dynamics generated by Eq.~\eqref{eq:appA_master} reduce to the deterministic equations~[Eq.\@~\eqref{eq:model_ode}] in the thermodynamic limit~$N \to \infty$ at fixed densities $\sigma_{n,\alpha}=N_{n,\alpha}/(V_\alpha\rho_0)$. 
This matching requires
\begin{subequations}
\label{eq:appA_rates}
\begin{align}
c^{\mathrm{nuc}}_{\alpha}     &= \frac{2\,\tau_\alpha\,\mu}{V_\alpha}
                              = \frac{2\,\tau_\alpha\,\eta\,\nu}{V_\alpha},
                              \label{eq:appA_rates_nuc}\\[3pt]
c^{\mathrm{grow}}_{\alpha}    &= \frac{\tau_\alpha\,\nu}{V_\alpha},
                              \label{eq:appA_rates_grow}\\[3pt]
c^{\mathrm{exch}}_{n,\alpha}  &= \frac{\mathcal{D}_n\,\nu\,\rho_0}{\phi_\alpha}.
                              \label{eq:appA_rates_exch}
\end{align}
\end{subequations}
The inverse-volume factor $V_\alpha^{-1}$ in Eqs.\@~\eqref{eq:appA_rates_nuc} and~\eqref{eq:appA_rates_grow} encodes the standard conversion between rate constants in density units (used in the mean-field model) and stochastic rate constants per molecular pair. 
Finally, the inverse volume-fraction scaling $\phi_\alpha^{-1}$ in the per-particle exchange rate~\eqref{eq:appA_rates_exch} reflects our parameterization of exchange through a single volume-normalized rate $\mathcal{D}_n$: a particle in the smaller compartment must escape proportionally more often to produce the same mean flux, so that the densities $\sigma_{n,\alpha}$---rather than the absolute counts $N_{n,\alpha}$---equilibrate between the two compartments in the absence of chemical reactions.

Equations~\eqref{eq:appA_propensities} and~\eqref{eq:appA_rates}, together with the master equation~\eqref{eq:appA_master}, fully specify the stochastic model underlying all numerical results presented in the main text.
Given the dimensionless mean-field parameters $(\eta,\tau_\alpha,\phi_\alpha,\mathcal{D}_n,S,N)$ defined in Sec.~\ref{sec:model}, and adopting the (arbitrary) convention $\nu\!=\!1$ that sets the unit of time, the simulation proceeds by sampling initial conditions---each of the $N_0=SN$ subunits is independently placed in the slow compartment with probability $\phi_s$ and in the fast compartment with probability $\phi_f=1-\phi_s$---and by iterating Gillespie updates of $\mathbf{N}(t)$ until no further reactions can occur.

\subsection{Choice of sample size}
\label{app:sample_size_criterion}

For stochastic simulations, we define the final yield as the average fraction of all subunits integrated into finished structures.
To determine a confidence interval around the reported final yield values for each data point we use a bootstrapping estimate with simple resampling~\citep{efronNonparametricEstimatesStandard1981}.
We dynamically choose the number of realizations (sample size) for each data point such that the width of the estimated 99\% confidence interval is less than $0.005$, with a minimum number of $100$ realizations.
We omit the estimated confidence intervals in our figures, since they are considerably smaller than the plot markers.
Due to their higher computational cost, the sample sizes for the spatially extended bulk-boundary simulations (Fig.\@~\ref{fig:2d_membrane}) are chosen such that the width of the 99\% confidence interval is at most $0.05$.

\section{Final yield for effectively decoupled compartments}
\label{app:decoupled_compartments_yield}

We initialize the simulations of the two-compartment system with a total number of ${N_0 = SN}$ subunits, the stoichiometric amount to form exactly $N$ target structures of size $S$.
For a given compartment volume ratio of $\phi_s = V_s / V$, we initially place each of the $N_0$ subunits with probability $\phi_s$ into the slow compartment and with probability $\phi_f = (1 - \phi_s)$ into the fast compartment.
Thus, the average initial number of subunits in the slow and fast compartment are $N_s = \phi_s N_0$ and $N_f = \phi_f N_0$, respectively.

For effectively decoupled high-yield compartments [$\mathcal{D} \to 0$; $\eta \ll \eta^*$; Fig.\@~\ref{fig:results_yield_catstrophe_2comp}(d)], these initial conditions lead to an average final yield that depends on the average number of subunits per compartment and the target size.
Let~$\hat N_\alpha$ be the actual number of initial subunits in each compartment for a single realization of the stochastic process, whose mean is given as $\langle \hat N_\alpha \rangle = N_\alpha = \phi_\alpha N_0$.
Without particle exchange, maximal yield is only possible if~$\hat N_\alpha$ is an integer multiple of the target size~$S$ in both compartments, i.e.,\@~$\hat N_\alpha = k_\alpha S$ with $k_\alpha \in \{0,1,\dots,N\}$ and~$k_f + k_s = N$.
Since our chosen probability distribution of initial subunit numbers~$\hat N_\alpha$ is binomial and thus strongly peaked around its mean, we can expect high yield when~$N_\alpha = k_\alpha S$.
Therefore, the final yield for vanishing exchange rates is highest if the compartment volumes are given by~$\phi_\alpha = k_\alpha/N$.
E.g., for $N = 3$, there are four possible volume ratios that split~$N$ into two integers and hence exhibit local maxima in the final yield: $\phi_s^* \in \{0, \frac{1}{3}, \frac{2}{3}, 1\}$ [Fig.\@~\ref{fig:results_yield_catstrophe_2comp}(d)].

The minimal final yield for decoupled high-yield compartments can be estimated by assuming all target structures to be completed, except one in each compartment, i.e., $Y_{\mathrm{min}} = (N-2)/N = 1 - 2/N$.
Both the local maxima and minima in the final yield at \(\phi_s^*\) can be seen in our simulations for $\mathcal{D} \to 0$ [Fig.~\ref{fig:results_yield_catstrophe_2comp}(d)].

\section{Hybrid two-stage approximation to delay-facilitated assembly}
\label{app:two_stage_simulations}

To remove the impact of stochasticity during the first stage of delay-facilitated assembly, we prepare the two-compartment system with the typical state after this stage (see Sec.~\ref{sec:origin_of_catastrophe}).
To clearly define the first assembly stage, we approximate the full system by a system without particle exchange ($\mathcal{D} = 0$) and no reactions in the slow compartment ($\tau = 0$).
Reactions therefore only occur in the isolated fast compartment, which initially contains~$\phi_f N_0$ subunits on average (see Appendix~\ref{app:decoupled_compartments_yield}).
In the mean-field limit this isolated compartment reaches an absorbing state with a fixed density per structure size~$\sigma_{n,f}$ and zero free subunits~($\sigma_{1,f} = 0$).
We obtain this state by solving Eq.~\eqref{eq:model_ode} with~$\mathcal{D} = 0$, $\mathcal{\tau} = 0$, and an initial density~$\rho_0 = N_0 /V$.

\begin{figure}[ht]
    \includegraphics[]{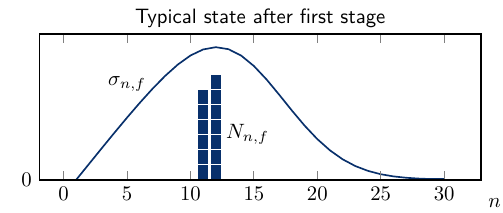}
    \caption{
    Example state in the fast compartment after the first assembly stage, without particle exchange ($\mathcal{D}=0$) for~$S = 30$, $N = 10$, $\phi_s = 0.5$, and~$\eta = 10\eta^*$.
    The mean-field density per structure size $\sigma_{n,f}$ is shown as solid line (arbitrary units) and the discretized state $N_{n,f}$ used for the hybrid simulations (see Sec.~\ref{sec:origin_of_catastrophe}) is represented by bars, here comprising six structures of size 11 and seven structures of size 12.
    }
    \label{fig:appendix:round_state}
\end{figure}

To use this final state as the initial condition for the second stage, re-introducing the corresponding values of~$\mathcal{D}$ and~$\tau$, we need to round the continuous particle densities to integer copy numbers per species~$N_{n,f}$.
We choose to round the densities such that the average number of structures is preserved and that the initial subunits are distributed as evenly as possible across that number of structures.
The average number of structures can be inferred directly from the densities in the fast compartment by multiplying with the compartment volume,
\begin{equation}
    N_f = \left\lceil \rho_0 V_f \sum_{n=2}^{S} \sigma_{n,f} \right \rceil \, ,
\end{equation}
where $\left\lceil \cdot \right\rceil$ denotes the ceiling function, ensuring that the discrete final state always contains at least one structure.
Using this scheme, we obtain, for example, that a system with~${S = 30}$, ${N = 10}$, ${\phi_s = 0.5}$, and~${\eta = 10\eta^*}$ typically contains $13$ structures, six of size $11$ and seven of size $12$ (Fig.~\ref{fig:appendix:round_state}).

The scheme described above only represents an approximation of the expected system state after the first assembly stage, which is justified as long as assembly in the slow compartment and particle exchange remain slow.
If either of these two processes is too fast~($\mathcal{D} \lesssim 1$ or $\tau \lesssim 1$), the distinction of the process into a first and second stage fails and we expect the hybrid simulation results to deviate from the original two-compartment simulations [$\mathcal{D} \to 1$ in Fig.~\ref{fig:mechanism}(b),(c)].

\section{Remaining stochasticity without structure exchange}
\label{app:stochasticity_without_structure_exchange}

Due to the timescale separation of delay-facilitated assembly, the two-compartment system first reaches a transient kinetic trap state in the fast compartment, before free subunits get slowly supplied to continue growing the nucleated structures, see Sec.~\ref{sec:origin_of_catastrophe}.
Any nucleation event during this first stage that raises the number of structures above the desired target number~$N$ prevents the system from completing all possible targets, regardless of the dynamics in the second stage.

If only subunits are able to exchange between compartments, we can estimate the maximal final yield from the statistics of the system state after the first stage, since no more structures are expected to nucleate during the second assembly stage in the delay-facilitated assembly regime.
For example, at~$\eta=10\eta^*$, $\phi_s = 0.6$, and~$N=1$, we find about 67\% of all individual realizations nucleate a single structure during the first stage, whereas the remaining realizations nucleate more.
Since~$N=1$ implies that either all target structures or none of them can assemble completely, the fraction of realizations with a single structure provides a direct estimate for the maximal final yield and agrees with our simulation results~[Fig.~\ref{fig:controlled_exchange}(a)].
For $N>1$ a correct estimate is more involved, as there are multiple combinations of initially nucleated structures and overall finished structures that have to be weighted appropriately to obtain the final yield.

\section{Hexagonal assembly model}
\label{app:hexagonal_assembly}

The assembly model of two-dimensional structures from hexagonal subunits (Fig.~\ref{fig:hexagons}) is inspired from Ref.~\citep{Gartner2024} and discussed in more detail therein.
Here, we only provide a short summary of the model and the stochastic simulations.

We assume that each subunit can attach to any other subunit or an existing structure at any available free edge with rate $\nu$.
Subunits can also detach from their binding partner with rate $\delta$ if they are only bound at a single edge [see Fig.~\ref{fig:hexagons}(a)].
All subunits bound with more than one edge are assumed to be stable on the relevant timescales of the process.
For simplicity, we assume the target structures to be periodic and contain $S=36$ subunits.
In our two-compartment system, every particle can exchange between compartments with rate $\mathcal{D}_n$, where $n$ is the number of subunits in the structure ($n=1$ denotes individual subunits).

The stochastic simulation results [Fig.\@~\ref{fig:hexagons}(b),(c)] are obtained from repeated simulations of the two-compartment system using the stochastic simulation algorithm (see Appendix~\ref{app:master_equation}).
In contrast to the linear ring model from the main text [Fig.~\ref{fig:recap_delay_facilitated_assembly}(a)], the state space of two-dimensional structures is practically too large to enumerate all possible structures and work with occupation numbers.
Instead, we represent individual structures explicitly and compute the relevant propensities dynamically (see Ref.~\citep{zenodo} for the implementation of both versions of the algorithms used in this study).

The mapping between the mean-field rates ($\nu$, $\delta$) and the microscopic rates for the Gillespie algorithm follows analogously to Eq.~\ref{eq:appA_reactions}, by matching the results of both simulations in the thermodynamic limit ($N \to \infty$).
In this limit, and for large target structures ($S \to \infty$), the assembly is well approximated by the following effective mean-field theory~\citep{Gartner2024}:
\begin{subequations}
\label{eq:appendix:hexagon_rate_equations}
    \begin{align}
        \partial_t\sigma_1 &= -m\bar\mu\sigma_1^m - \bar\nu\sigma_1^\gamma \sum_{n=m}^{S-1}f_n \sigma_n \\
        \partial_t\sigma_m &= \bar\mu\sigma_1^m - \bar\nu\sigma_1^\gamma f_m\sigma_m \\
        \partial_t\sigma_{m<n<S} &= \bar\nu\sigma_1^\gamma(f_{n-1}\sigma_{n-1} - f_n\sigma_n) \\
        \partial_t\sigma_S &= \bar\nu\sigma_1^\gamma f_{S-1}\sigma_{S-1}
    \end{align}
\end{subequations}
The effective rate equations~\eqref{eq:appendix:hexagon_rate_equations} are structurally similar to Eq.\@~\eqref{eq:model_ode} and can be extended by the same exchange terms.
Instead of microscopic nucleation and growth rates, Eq.~\ref{eq:appendix:hexagon_rate_equations} contains the effective nucleation rate~${\bar\mu = \nu(\nu\rho_0/\delta)^{m-1}}$ and effective growth rate~${\bar\nu = \nu(\nu\rho_0/\delta)^{\gamma-1}}$.
The remaining parameters are the nucleation size $m$, the attachment order $\gamma$, and a combinatorial prefactor $f_n$ which scales with structure size.
For hexagonal subunits, the appropriate parameters are $m = 3$, $\gamma = 1$ and $f_n = 2.3 \, n^{1/2}$, but the same set of equations can also describe structures with different subunit morphologies~\citep{Gartner2024}.
The dimensionless nucleation-to-growth ratio determining the final yield of the hexagonal assemblies is given by $\eta = \nu \rho_0/\delta$.

The small systematic discrepancy between the mean-field results and the stochastic simulations for large $N$ [compare $N=\infty$ and $N=100$ in  Fig.~\ref{fig:hexagons}(b)] stems from the finite target structure size and vanishes in the limit $S \to \infty$.

\section{2D cytosol-membrane geometry}
\label{app:cytosol_membrane_assembly}

The spatially extended system representing a cytosolic bulk with a reactive membrane as boundary (Fig.~\ref{fig:2d_membrane}) features the same reaction dynamics described in the main text [Fig.~\ref{fig:recap_delay_facilitated_assembly}(a)], with the cytosol representing the slow compartment and the membrane the fast compartment.
In contrast to the conceptual two-compartment model, however, the bulk-boundary system can resolve spatial density gradients and fluctuations.
Particle exchange is no longer determined by a single rate, but through diffusion in the bulk and on the boundary with mean-field rates $D$ and $\bar D$ (a bar on top of any quantity signifies its value on the boundary), and structure-size independent particle attachment an detachment rates $k_a$ and $k_d$.
The corresponding mean-field equations for the subunit density in a two-dimensional circular domain of radius~$R$ [used for the deterministic data in Fig.\@~\ref{fig:2d_membrane}] are derived in Ref.\@~\cite{angerpointnerDelayfacilitatedSelfassemblyCompartmentalized2025} and are given by
\begin{subequations}
\label{eq:appendix:bulk-boundary-mean-field}
    \begin{align}
        \partial_t \sigma_1(r) &= 
        \mathcal{D} \partial_r^2 \sigma_1(r) - \tau \left[ 2 \eta \sigma_1(r)^2 + \sigma_1(r)\sum_{n=2}^{S-1} \sigma_n(r) \right]
        \\
        \partial_t \bar\sigma_1 &=
         \mathcal{D}_d \left[ \sigma_1(1) - \bar\sigma_1 \right]
        - 2 \bar\eta \bar\sigma_1^2 - \bar\sigma_1 \sum_{n=2}^{S-1} \bar\sigma_n
        \\
        \partial_r \sigma_1(1) &= 
        \mathcal{D}_a \left[\bar\sigma_1 - \sigma_1(1)\right] \, , ~~ r\in[0,1] \, .
    \end{align}
\end{subequations}
Equations for the structure densities $\sigma_{n>1}$ follow analogously after replacing the reaction terms according to Eq.\@~\eqref{eq:model_ode}.
The dimensionless parameters are given by
\begin{subequations}
\label{eq:app:2d_non_dim_quantities}
\begin{gather}
    \eta = \frac{\mu}{\nu} \, , \ 
    \bar\eta = \frac{\bar\mu}{\bar\nu} \, , \ 
    \tau = \frac{\nu \rho_0}{\bar\nu \bar\rho_0} \, , \\
    \mathcal{D} = \frac{D}{\bar\nu \bar\rho_0 R^2} \, , \
    \mathcal{D}_d = \frac{k_d}{\bar\nu \bar\rho_0} \, , \
    \mathcal{D}_a = \frac{k_a R}{D} \, ,
\end{gather}
\end{subequations}
and the dimensionless volumes, i.e., initial subunit numbers, are given by
\begin{equation}
    \label{eq:app:2d_non_dim_vols}
    V_\mathrm{bulk} = \pi R^2 \rho_0 \, , \ 
    V_\mathrm{bound} = 2 \pi R \bar\rho_0 \, , \
    \phi_s = \frac{R}{R + 2 \bar\rho_0/\rho_0} \, .
\end{equation}
Eqs.\@~\eqref{eq:appendix:bulk-boundary-mean-field} contain no boundary diffusion term or angular gradients, since we assume a homogeneous initial density and thus radial symmetry on the mean-field level.

In the stochastic particle-based simulations, for simplicity, we model the cytosolic bulk as a 2D square-shaped domain of side length $R$ [Fig.~\ref{fig:2d_membrane}(a)].
For target numbers $N=1,10,100$, we discretize the spatial domain with a lattice of $20 \times 20$, $30 \times 30$, and $40 \times 40$ sites such that in each case the impact of changing the number of lattice sites is minimal.
The boundary sites surrounding the lattice represent the membrane.
As for the two-compartment system in Appendix~\ref{app:master_equation}, the microscopic rate constants for diffusion, attachment and detachment can be mapped to the mean-field rates via the master equation.

The initial total number of subunits is fixed by the target number $N$ and for the data shown in Fig.~\ref{fig:2d_membrane} the bulk size and the microscopic rate constant are chosen such that, for each data point, they correspond to non-dimensionalized mean-field quantities Eqs.\@~\eqref{eq:app:2d_non_dim_quantities}\nobreakdash--\eqref{eq:app:2d_non_dim_vols} with values $\eta=10\eta^*$, $\tau=10^{-6}$, $\phi_s=0.6$, $\mathcal{D}_d = 10\mathcal{D}$ (for every value of $\mathcal{D}$), and $\bar{\mathcal{D}}=\bar D / (\bar\nu\bar\rho_0R^2) = 100$.
This choice of parameters guarantees that particle exchange is limited by bulk diffusion (since attachment/detachment-limited exchange renders the system equivalent to the two-compartment system) and that the assembly on the boundary (the high-reactivity domain) is not limited by membrane diffusion.
For a fixed target number $N$, the bulk exchange parameter $\mathcal{D}$ is then varied by changing the bulk diffusion constant and keeping everything else fixed.
While keeping the boundary diffusion fixed at a large value is the cleanest way to extract the effects of stochastic fluctuations without superimposing them with deterministic effects coming from a diffusion-limited assembly on the boundary, it can result in significantly faster boundary diffusion compared to bulk diffusion, which is biologically not very realistic.

\section{Size-dependent exchange at large target numbers}
\label{app:size-dependent_exchange_mean_field}

\begin{figure}[ht]
    \includegraphics[]{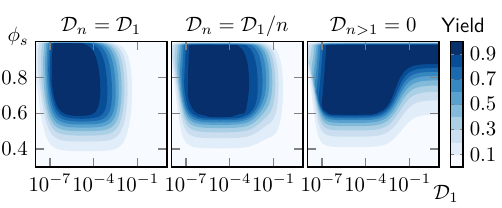}
    \caption{
    Delay-facilitated assembly unaffected by different structure-size dependent exchange rates in the mean-field limit.
    Final yield as a function of the subunit exchange parameter~$\mathcal{D}_1$ and relative slow compartment volume $\phi_s$ for three different functional dependencies of the exchange parameter $\mathcal{D}_n$ on the structure size $n$: $\mathcal{D}_n = \mathcal{D}_1$ (left), $\mathcal{D}_n = \mathcal{D}_1/n$ (middle), and $\mathcal{D}_{n>1} = 0$ (right).
    For two coupled low-yield compartments ($\eta = 10\eta^*$) and relative compartment reactivity $\tau=10^{-6}$, all choices result in the same region of improved final yield at an intermediate subunit exchange parameter.
    }
    \label{fig:appendix:size-dependent_exchange_mean_field}
\end{figure}

Whether or not exchange rates depend on structure size in the mean-field two-compartment model [Eq.\@~\eqref{eq:model_ode}], affects the delay-facilitated assembly mechanism only minimally.
We perform mean-field simulations of Eq.~\eqref{eq:model_ode} using the same functional dependencies of the exchange rates $\mathcal{D}_n$ on structure size $n$ as described in the main text (see Sec.\@~\ref{sec:control-exchange}), namely equal exchange for all subunits and structures ($\mathcal{D}_n = \mathcal{D}_1$, left plot in Fig.\@~\ref{fig:appendix:size-dependent_exchange_mean_field}), proportionally decreasing exchange rates with size ($\mathcal{D}_n = \mathcal{D}_1 / n$, center plot in Fig.\@~\ref{fig:appendix:size-dependent_exchange_mean_field}), and exclusive exchange for subunits ($\mathcal{D}_{n>1} = 0$, right plot in Fig.\@~\ref{fig:appendix:size-dependent_exchange_mean_field}).
All three cases exhibit delay-facilitated self-assembly in the same parameter regime (Fig.\@~\ref{fig:appendix:size-dependent_exchange_mean_field}).

The only qualitative difference between exclusive exchange of subunits and the remaining cases is that for fast subunit exchange ($\mathcal{D}_1 \rightarrow \infty$) the final yield is not zero, but approaches a constant, which depends on the relative compartment volumes (Fig.\@~\ref{fig:appendix:size-dependent_exchange_mean_field}).
The reason for this non-vanishing yield is that since only subunits exchange, the system does not approach well-mixed dynamics for $\mathcal{D}_1 \rightarrow \infty$.
Instead, the structure density in the fast compartment remains high throughout and is increased by a factor of $1/\phi_f$ compared to the well-mixed case.
This in turns lowers the effective nucleation-to-growth ratio by a factor of $1/\phi_f$, allowing for higher yields for decreasing fast-compartment size.
However, in the regime of delay-facilitated assembly (intermediate exchange parameters) this distinction does not matter and all different exchange dynamics essentially provide the same final yields.

\section{Stochastic yield catastrophe independent of target size}
\label{app:diff-target-sizes}

\begin{figure}[ht]
    \includegraphics[]{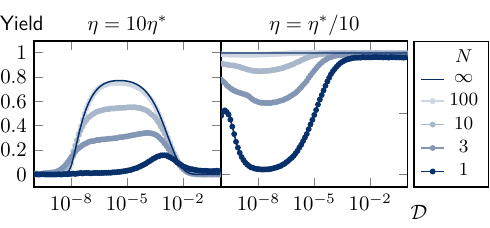}
    \caption{
    Delay-induced stochastic yield catastrophe for targets of size~$S=120$.
    Final yield as a function of the size-independent exchange parameter $\mathcal{D} \equiv \mathcal{D}_n$ for two coupled low-yield ($\eta=10\eta^*$; left panel) and high-yield ($\eta=\eta^*/10$; right panel) compartments for different target numbers $N$.
    The relative compartment volume and relative compartment reactivity are fixed to $\phi_s=0.6$ and $\tau=10^{-6}$.
    The same delay-induced stochastic yield catastrophe as for structure of size $S=30$ emerges.
    }
    \label{fig:appendix:effect_doesnt_depend_on_structure_size}
\end{figure}

The well-mixed assembly model [Fig.~\ref{fig:recap_delay_facilitated_assembly}(a)] features an optimal nucleation-to-growth ratio~$\eta^*$ that depends on the structure's size~$S$.
Intuitively, larger structures require more growth reactions in sequence to get completed, which in turn requires slower nucleation to ensure every nucleated structure gets completed before the next nucleation occurs.
For an in-depth discussion and derivation of the asymptotic structure size dependence of~$\eta^* \sim S^{-2}$ and the fact that this class of assembly models exhibits scaling behavior with~$S$ in all relevant observables, we refer to Ref.~\citep{gartnerStochasticYieldCatastrophes2020}.
Here, we only show that there is no qualitative change in behavior in both delay-facilitated assembly and the delay-induced stochastic yield catastrophe for structures of different size [compare Fig.\@~\ref{fig:appendix:effect_doesnt_depend_on_structure_size} with Fig.\@~\ref{fig:results_yield_catstrophe_2comp}(a)\nobreakdash--(b)].

\newpage

\bibliography{references}

@article{angerpointnerDelayfacilitatedSelfassemblyCompartmentalized2025, 
    year = {2025}, 
    title = {{Delay-facilitated self-assembly in compartmentalized systems}}, 
    author = {Angerpointner, Severin and Swiderski, Richard and Frey, Erwin}, 
    journal = {Proceedings of the National Academy of Sciences}, 
    issn = {0027-8424}, 
    doi = {10.1073/pnas.2515123122}, 
    pmid = {41289406}, 
    pmcid = {PMC12685093}, 
    eprint = {2508.03383}, 
    pages = {e2515123122}, 
    number = {48}, 
    volume = {122}
}

@article{gartnerStochasticYieldCatastrophes2020,
  title = {Stochastic yield catastrophes and robustness in self-assembly},
  author = {Gartner, Florian M and Graf, Isabella R and Wilke, Patrick and Geiger, Philipp M and Frey, Erwin},
  year = 2020,
  month = feb,
  journal = {eLife},
  volume = {9},
  pages = {e51020},
  issn = {2050-084X},
  doi = {10.7554/eLife.51020},
  url = {https://elifesciences.org/articles/51020},
  urldate = {2021-11-06},
  langid = {english},
}

@article{Zlotnick2011VirusCapsids, 
year = {2011}, 
title = {{Virus assembly, allostery and antivirals}}, 
author = {Zlotnick, Adam and Mukhopadhyay, Suchetana}, 
journal = {Trends in Microbiology}, 
issn = {0966-842X}, 
doi = {10.1016/j.tim.2010.11.003}, 
pmid = {21163649}, 
pmcid = {PMC3026312}, 
pages = {14--23}, 
number = {1}, 
volume = {19}, 
}

@article{Perlmutter2015VirusCapsids, 
year = {2015}, 
title = {{Mechanisms of virus assembly}}, 
author = {Perlmutter, Jason D. and Hagan, Michael F.}, 
journal = {Annual Review of Physical Chemistry}, 
issn = {0066-426X}, 
doi = {10.1146/annurev-physchem-040214-121637}, 
pmid = {25532951}, 
pmcid = {PMC4382372}, 
eprint = {1407.3856}, 
pages = {1--23}, 
number = {1}, 
volume = {66}, 
}

@article{Basler2018RibosomeAssemblyEukaryotes, 
year = {2018}, 
title = {{Eukaryotic ribosome assembly}}, 
author = {Baßler, Jochen and Hurt, Ed}, 
journal = {Annual Review of Biochemistry}, 
issn = {0066-4154}, 
doi = {10.1146/annurev-biochem-013118-110817}, 
pmid = {30566372}, 
pages = {1--26}, 
number = {1}, 
volume = {88}
}

@article{Shajani2011RibosomeBacterial, 
year = {2011}, 
title = {{Assembly of bacterial ribosomes}}, 
author = {Shajani, Zahra and Sykes, Michael T. and Williamson, James R.}, 
journal = {Annual Review of Biochemistry}, 
issn = {0066-4154}, 
doi = {10.1146/annurev-biochem-062608-160432}, 
pmid = {21529161}, 
pages = {501--526}, 
number = {1}, 
volume = {80}
}

@article{Kerfeld2018BMCreview, 
year = {2018}, 
title = {{Bacterial microcompartments}}, 
author = {Kerfeld, Cheryl A. and Aussignargues, Clement and Zarzycki, Jan and Cai, Fei and Sutter, Markus}, 
journal = {Nature Reviews Microbiology}, 
issn = {1740-1526}, 
doi = {10.1038/nrmicro.2018.10}, 
pmid = {29503457}, 
pmcid = {PMC6022854}, 
pages = {277--290}, 
number = {5}, 
volume = {16},
}

@article{ChevanceFlagellumOverview, 
year = {2008}, 
title = {{Coordinating assembly of a bacterial macromolecular machine}}, 
author = {Chevance, Fabienne F. V. and Hughes, Kelly T.}, 
journal = {Nature Reviews Microbiology}, 
issn = {1740-1526}, 
doi = {10.1038/nrmicro1887}, 
pmid = {18483484}, 
pmcid = {PMC5963726}, 
pages = {455--465}, 
number = {6}, 
volume = {6}, 
}

@article{siglProgrammableIcosahedralShell2021, 
year = {2021}, 
title = {{Programmable icosahedral shell system for virus trapping}}, 
author = {Sigl, Christian and Willner, Elena M. and Engelen, Wouter and Kretzmann, Jessica A. and Sachenbacher, Ken and Liedl, Anna and Kolbe, Fenna and Wilsch, Florian and Aghvami, S. Ali and Protzer, Ulrike and Hagan, Michael F. and Fraden, Seth and Dietz, Hendrik}, 
journal = {Nature Materials}, 
issn = {1476-1122}, 
doi = {10.1038/s41563-021-01020-4}, 
pmid = {34127822}, 
pmcid = {PMC7611604}, 
pages = {1281--1289}, 
number = {9}, 
volume = {20}
}

@article{monferrerDNAOrigamiTraps2023,
  title = {{{DNA}} origami traps for large viruses},
  author = {Monferrer, Alba and Kohler, Fabian and Sigl, Christian and Schachtner, Michael and Peterhoff, David and Asbach, Benedikt and Wagner, Ralf and Dietz, Hendrik},
  year = {2023},
  month = jan,
  journal = {Cell Reports Physical Science},
  volume = {4},
  number = {1},
  pages = {101237},
  issn = {26663864},
  doi = {10.1016/j.xcrp.2022.101237},
  url = {https://linkinghub.elsevier.com/retrieve/pii/S266638642200563X},
  urldate = {2025-06-12},
  langid = {english},
}

@article{khmelinskaia_structure-based_2021,
	title = {Structure-based design of novel polyhedral protein nanomaterials},
	volume = {61},
	issn = {13695274},
	url = {https://linkinghub.elsevier.com/retrieve/pii/S1369527421000382},
	doi = {10.1016/j.mib.2021.03.003},
	pages = {51--57},
	journaltitle = {Current Opinion in Microbiology},
	journal = {Current Opinion in Microbiology},
	shortjournal = {Current Opinion in Microbiology},
	author = {Khmelinskaia, Alena and Wargacki, Adam and King, Neil P},
	urldate = {2024-01-29},
	date = {2021-06},
	year = {2021},
	langid = {english},
}

@article{jiangDNAOrigamiMolecular2024,
  title = {{{DNA origami}}: {{from molecular folding art}} to {{drug delivery technology}}},
  shorttitle = {{{DNA Origami}}},
  author = {Jiang, Qiao and Shang, Yingxu and Xie, Yiming and Ding, Baoquan},
  year = {2024},
  month = may,
  journal = {Advanced Materials},
  volume = {36},
  number = {22},
  pages = {2301035},
  issn = {0935-9648, 1521-4095},
  doi = {10.1002/adma.202301035},
  url = {https://onlinelibrary.wiley.com/doi/10.1002/adma.202301035},
  urldate = {2025-06-12},
  langid = {english},
}

@article{zlotnickTheoreticalModelSuccessfully1999,
  title = {A {{theoretical model successfully identifies features}} of {{hepatitis B virus capsid assembly}}},
  author = {Zlotnick, Adam and Johnson, Jennifer M. and Wingfield, Paul W. and Stahl, Stephen J. and Endres, Dan},
  year = {1999},
  month = nov,
  journal = {Biochemistry},
  volume = {38},
  number = {44},
  pages = {14644--14652},
  issn = {0006-2960, 1520-4995},
  doi = {10.1021/bi991611a},
  url = {https://pubs.acs.org/doi/10.1021/bi991611a},
  urldate = {2025-06-10},
  langid = {english},
}

@article{morozov_assembly_2009,
	title = {Assembly of viruses and the pseudo-law of mass action},
	volume = {131},
	issn = {0021-9606, 1089-7690},
	url = {https://pubs.aip.org/jcp/article/131/15/155101/317050/Assembly-of-viruses-and-the-pseudo-law-of-mass},
	doi = {10.1063/1.3212694},
	pages = {155101},
	number = {15},
	journaltitle = {The Journal of Chemical Physics},
	journal = {The Journal of Chemical Physics},
	author = {Morozov, Alexander Yu. and Bruinsma, Robijn F. and Rudnick, Joseph},
	urldate = {2025-04-27},
	date = {2009-10-21},
	year = {2009},
	langid = {english},
}

@article{hagan_understanding_2010,
	title = {Understanding the concentration dependence of viral capsid assembly kinetics—the origin of the lag time and identifying the critical nucleus size},
	volume = {98},
	issn = {00063495},
	url = {https://linkinghub.elsevier.com/retrieve/pii/S0006349509017524},
	doi = {10.1016/j.bpj.2009.11.023},
	pages = {1065--1074},
	number = {6},
	journaltitle = {Biophysical Journal},
	journal = {Biophysical Journal},
	shortjournal = {Biophysical Journal},
	author = {Hagan, Michael F. and Elrad, Oren M.},
	urldate = {2025-02-25},
	date = {2010-03},
	year = {2010},
	langid = {english},
}

@article{haganMechanismsKineticTrapping2011,
  title = {Mechanisms of kinetic trapping in self-assembly and phase transformation},
  author = {Hagan, Michael F. and Elrad, Oren M. and Jack, Robert L.},
  year = {2011},
  month = sep,
  journal = {The Journal of Chemical Physics},
  volume = {135},
  number = {10},
  pages = {104115},
  issn = {0021-9606, 1089-7690},
  doi = {10.1063/1.3635775},
  url = {https://pubs.aip.org/jcp/article/135/10/104115/983390/Mechanisms-of-kinetic-trapping-in-self-assembly},
  urldate = {2024-01-09},
  langid = {english},
}

@article{keThreeDimensionalStructuresSelfAssembled2012,
  title = {Three-{{dimensional structures self-assembled}} from {{DNA bricks}}},
  author = {Ke, Yonggang and Ong, Luvena L. and Shih, William M. and Yin, Peng},
  year = {2012},
  month = nov,
  journal = {Science},
  volume = {338},
  number = {6111},
  pages = {1177--1183},
  issn = {0036-8075, 1095-9203},
  doi = {10.1126/science.1227268},
  url = {https://www.science.org/doi/10.1126/science.1227268},
  urldate = {2025-06-10},
  langid = {english},
}

@article{weiComplexShapesSelfassembled2012,
  title = {Complex shapes self-assembled from single-stranded {{DNA}} tiles},
  author = {Wei, Bryan and Dai, Mingjie and Yin, Peng},
  year = {2012},
  month = may,
  journal = {Nature},
  volume = {485},
  number = {7400},
  pages = {623--626},
  issn = {0028-0836, 1476-4687},
  doi = {10.1038/nature11075},
  url = {https://www.nature.com/articles/nature11075},
  urldate = {2025-06-10},
  copyright = {http://www.springer.com/tdm},
  langid = {english},
}

@article{reinhardtNumericalEvidenceNucleated2014,
  title = {Numerical {{evidence}} for {{nucleated self-assembly}} of {{DNA brick structures}}},
  author = {Reinhardt, Aleks and Frenkel, Daan},
  year = {2014},
  month = jun,
  journal = {Physical Review Letters},
  volume = {112},
  number = {23},
  pages = {238103},
  issn = {0031-9007, 1079-7114},
  doi = {10.1103/PhysRevLett.112.238103},
  url = {https://link.aps.org/doi/10.1103/PhysRevLett.112.238103},
  urldate = {2025-06-10},
  copyright = {http://link.aps.org/licenses/aps-default-license},
  langid = {english},
}

@article{jacobsRationalDesignSelfassembly2015,
  title = {Rational design of self-assembly pathways for complex multicomponent structures},
  author = {Jacobs, William M. and Reinhardt, Aleks and Frenkel, Daan},
  year = {2015},
  month = may,
  journal = {Proceedings of the National Academy of Sciences},
  volume = {112},
  number = {20},
  pages = {6313--6318},
  issn = {0027-8424, 1091-6490},
  doi = {10.1073/pnas.1502210112},
  url = {https://pnas.org/doi/full/10.1073/pnas.1502210112},
  urldate = {2025-06-10},
  langid = {english},
}

@article{pintoDesignStrategiesSelfassembly2023,
  title = {Design strategies for the self-assembly of polyhedral shells},
  author = {Pinto, Diogo E. P. and {\v S}ulc, Petr and Sciortino, Francesco and Russo, John},
  year = {2023},
  month = apr,
  journal = {Proceedings of the National Academy of Sciences},
  volume = {120},
  number = {16},
  pages = {e2219458120},
  issn = {0027-8424, 1091-6490},
  doi = {10.1073/pnas.2219458120},
  url = {https://pnas.org/doi/10.1073/pnas.2219458120},
  urldate = {2025-06-11},
  langid = {english},
}

@article{rovigattiSimpleSolutionProblem2022,
  title = {A simple solution to the problem of self-assembling cubic diamond crystals},
  author = {Rovigatti, Lorenzo and Russo, John and Romano, Flavio and Matthies, Michael and Kroc, Luk{\'a}{\v s} and {\v S}ulc, Petr},
  year = {2022},
  journal = {Nanoscale},
  volume = {14},
  number = {38},
  pages = {14268--14275},
  issn = {2040-3364, 2040-3372},
  doi = {10.1039/D2NR03533B},
  url = {https://xlink.rsc.org/?DOI=D2NR03533B},
  urldate = {2025-06-11},
  langid = {english},
}

@article{haganSelfassemblyCoupledLiquidliquid2023,
	title = {Self-assembly coupled to liquid-liquid phase separation},
	volume = {19},
	issn = {1553-7358},
	url = {https://dx.plos.org/10.1371/journal.pcbi.1010652},
	doi = {10.1371/journal.pcbi.1010652},
	pages = {e1010652},
	number = {5},
	journal = {{PLOS} Computational Biology},
	shortjournal = {{PLoS} Comput Biol},
	author = {Hagan, Michael F. and Mohajerani, Farzaneh},
	editor = {Zhou, Huan-Xiang},
	urldate = {2024-06-19},
	date = {2023-05-15},
	year = {2023},
	langid = {english},
}

@article{laha_chemical_2024,
    title = {Chemical {{reactions}} regulated by {{phase-separated condensates}}},
    author = {Laha, Sudarshana and Bauermann, Jonathan and J{\"u}licher, Frank and Michaels, Thomas C. T. and Weber, Christoph A.},
    year = {2024},
    journal = {Physical Review Research}, 
    doi = {10.1103/physrevresearch.6.043092}, 
    pages = {043092}, 
    number = {4}, 
    volume = {6}
}

@article{bartolucci_interplay_2024, 
    title={The interplay between biomolecular assembly and phase separation}, 
    journal = {eLife},  
    author={Bartolucci, Giacomo and Haugerud, Ivar S. and Michaels, Thomas C.T. and Weber, Christoph A.}, 
    year={2026}, 
    volume = {13},
    doi = {10.7554/elife.93003.3}, 
    
}

@article{frechetteComputerSimulationsShow2025,
  title = {Computer simulations show that liquid-liquid phase separation enhances self-assembly},
  author = {Frechette, Layne B. and Sundararajan, Naren and Caballero, Fernando and Trubiano, Anthony and Hagan, Michael F.},
  year = {2025},
  author = {Frechette, Layne B. and Sundararajan, Naren and Caballero, Fernando and Trubiano, Anthony and Hagan, Michael F.}, 
journal = {ACS Nano}, 
issn = {1936-0851}, 
doi = {10.1021/acsnano.5c08120}, 
pmid = {40782375}, 
pmcid = {PMC12392740}, 
pages = {30275--30291}, 
number = {33}, 
volume = {19}
}

@book{phillipsPhysicalBiologyCell2013odetoecoli,
  author = {Phillips, Rob},
  title = {Physical biology of the cell},
  year = {2013},
  pages = {39-49,121-122},
  edition = {2},
  doi = {https://doi.org/10.1201/9781134111589},
  publisher = {Garland Science},
  address = {London: New York, NY},
  isbn = {978-0-8153-4450-6},
  lccn = {QH505 .P455 2013},
}

@article{hillLifeCycleCyanobacterial2020,
  title = {Life cycle of a cyanobacterial carboxysome},
  author = {Hill, Nicholas C. and Tay, Jian Wei and Altus, Sabina and Bortz, David M. and Cameron, Jeffrey C.},
  year = 2020,
  month = may,
  journal = {Science Advances},
  volume = {6},
  number = {19},
  pages = {eaba1269},
  issn = {2375-2548},
  doi = {10.1126/sciadv.aba1269},
  urldate = {2026-05-06},
  langid = {english}
}

@article{schaefferStochasticEmergenceTwo2022,
  title = {Stochastic {{emergence}} of {{two distinct self-replicators}} from a {{dynamic combinatorial library}}},
  author = {Schaeffer, Ga{\"e}l and Eleveld, Marcel J. and Ottel{\'e}, Jim and Kroon, Peter C. and Frederix, Pim W. J. M. and Yang, Shuo and Otto, Sijbren},
  year = 2022,
  month = apr,
  journal = {Journal of the American Chemical Society},
  volume = {144},
  number = {14},
  pages = {6291--6297},
  issn = {0002-7863, 1520-5126},
  doi = {10.1021/jacs.1c12591},
  urldate = {2025-10-23},
  copyright = {https://creativecommons.org/licenses/by/4.0/},
  langid = {english}
}

@article{tiwariStochasticLagTime2016,
  title = {Stochastic lag time in nucleated linear self-assembly},
  author = {Tiwari, Nitin S. and Van Der Schoot, Paul},
  year = 2016,
  month = jun,
  journal = {The Journal of Chemical Physics},
  volume = {144},
  number = {23},
  pages = {235101},
  issn = {0021-9606, 1089-7690},
  doi = {10.1063/1.4953850},
  urldate = {2026-05-10},
  langid = {english}
}

@article{michaelsFluctuationsKineticsLinear2016,
  title = {Fluctuations in the {{kinetics}} of {{linear protein self-assembly}}},
  author = {Michaels, Thomas C. T. and Dear, Alexander J. and Kirkegaard, Julius B. and Saar, Kadi L. and Weitz, David A. and Knowles, Tuomas P. J.},
  year = 2016,
  month = jun,
  journal = {Physical Review Letters},
  volume = {116},
  number = {25},
  pages = {258103},
  issn = {0031-9007, 1079-7114},
  doi = {10.1103/PhysRevLett.116.258103},
  urldate = {2026-05-10},
  copyright = {http://link.aps.org/licenses/aps-default-license},
  langid = {english}
}

@article{davisInitialConditionStochastic2016,
  title = {Initial condition of stochastic self-assembly},
  author = {Davis, Jason K. and Sindi, Suzanne S.},
  year = 2016,
  month = feb,
  journal = {Physical Review E},
  volume = {93},
  number = {2},
  pages = {022109},
  issn = {2470-0045, 2470-0053},
  doi = {10.1103/PhysRevE.93.022109},
  urldate = {2025-11-04},
  copyright = {http://link.aps.org/licenses/aps-default-license},
  langid = {english}
}

@article{Gillespie1976,
title = {A general method for numerically simulating the stochastic time evolution of coupled chemical reactions},
journal = {Journal of Computational Physics},
volume = {22},
number = {4},
pages = {403-434},
year = {1976},
issn = {0021-9991},
doi = {https://doi.org/10.1016/0021-9991(76)90041-3},
url = {https://www.sciencedirect.com/science/article/pii/0021999176900413},
author = {Daniel T Gillespie}
}

@article{Gillespie.2007, 
year = {2007}, 
title = {{Stochastic simulation of chemical kinetics}}, 
author = {Gillespie, Daniel T.}, 
journal = {Annual Review of Physical Chemistry}, 
issn = {0066-426X}, 
doi = {10.1146/annurev.physchem.58.032806.104637}, 
pmid = {17037977}, 
pages = {35--55}, 
number = {1}, 
volume = {58}
}

@article{Gartner2024, 
year = {2024}, 
title = {{Design principles for fast and efficient self-assembly processes}}, 
author = {Gartner, Florian M. and Frey, Erwin}, 
journal = {Physical Review X}, 
doi = {10.1103/physrevx.14.021004}, 
pages = {021004}, 
number = {2}, 
volume = {14},
}

@article{Kojima2020, 
year = {2020}, 
title = {{Regulation of the single polar flagellar biogenesis}}, 
author = {Kojima, Seiji and Terashima, Hiroyuki and Homma, Michio}, 
journal = {Biomolecules}, 
doi = {10.3390/biom10040533}, 
pmid = {32244780}, 
pmcid = {PMC7226244}, 
pages = {533}, 
number = {4}, 
volume = {10}
}

@article{Bange2025, 
year = {2025}, 
title = {{Where and how many: evolutionary diversification of a molecular switch regulating flagellar patterns}}, 
author = {Bange, Gert and Hochberg, Georg and Thormann, Kai and Dornes, Anita}, 
journal = {Journal of Bacteriology}, 
issn = {0021-9193}, 
doi = {10.1128/jb.00329-25}, 
pmid = {41358728}, 
pmcid = {PMC12826062}, 
pages = {e00329--25}, 
number = {1}, 
volume = {208}
}

@article{Nakamura2024, 
year = {2024}, 
title = {{Structure and dynamics of the bacterial flagellar motor complex}}, 
author = {Nakamura, Shuichi and Minamino, Tohru}, 
journal = {Biomolecules}, 
doi = {10.3390/biom14121488}, 
pmid = {39766194}, 
pmcid = {PMC11673145}, 
pages = {1488}, 
number = {12}, 
volume = {14}
}

@article{delbruckBurstSizeDistribution1945,
  title = {The {{burst size distribution}} in the {{growth}} of {{bacterial viruses}} ({{bacteriophages}})},
  author = {Delbr{\"u}ck, M.},
  year = 1945,
  month = aug,
  journal = {Journal of Bacteriology},
  volume = {50},
  number = {2},
  pages = {131--135},
  issn = {0021-9193, 1098-5530},
  doi = {10.1128/jb.50.2.131-135.1945},
  urldate = {2026-05-27},
  langid = {english}
}

@article{paradaViralBurstSize2006,
  title = {Viral burst size of heterotrophic prokaryotes in aquatic systems},
  author = {Parada, Ver{\'o}nica and Herndl, Gerhard J. and Weinbauer, Markus G.},
  year = 2006,
  month = jun,
  journal = {Journal of the Marine Biological Association of the United Kingdom},
  volume = {86},
  number = {3},
  pages = {613--621},
  issn = {0025-3154, 1469-7769},
  doi = {10.1017/S002531540601352X},
  urldate = {2026-05-27},
  copyright = {https://www.cambridge.org/core/terms},
  langid = {english}
}

@article{brussaardViralControlPhytoplankton2004,
  title = {Viral {{control}} of {{phytoplankton populations}}---a {{review}}},
  author = {Brussaard, Corina P. D.},
  year = 2004,
  month = mar,
  journal = {Journal of Eukaryotic Microbiology},
  volume = {51},
  number = {2},
  pages = {125--138},
  issn = {1066-5234, 1550-7408},
  doi = {10.1111/j.1550-7408.2004.tb00537.x},
  urldate = {2026-05-27},
  langid = {english}
}

@article{chenDeterminationVirusBurst2007,
  title = {Determination of virus burst size {\emph{in vivo}} using a single-cycle {{SIV}} in Rhesus Macaques},
  author = {Chen, Hannah Yuan and Di Mascio, Michele and Perelson, Alan S. and Ho, David D. and Zhang, Linqi},
  year = 2007,
  month = nov,
  journal = {Proceedings of the National Academy of Sciences},
  volume = {104},
  number = {48},
  pages = {19079--19084},
  issn = {0027-8424, 1091-6490},
  doi = {10.1073/pnas.0707449104},
  urldate = {2026-05-27},
  langid = {english}
}

@article{Chowdhury2015SelectiveProteinShells, 
year = {2015}, 
title = {{Selective molecular transport through the protein shell of a bacterial microcompartment organelle}}, 
author = {Chowdhury, Chiranjit and Chun, Sunny and Pang, Allan and Sawaya, Michael R. and Sinha, Sharmistha and Yeates, Todd O. and Bobik, Thomas A.}, 
journal = {Proceedings of the National Academy of Sciences}, 
issn = {0027-8424}, 
doi = {10.1073/pnas.1423672112}, 
pmid = {25713376}, 
pmcid = {PMC4364225}, 
pages = {2990--2995}, 
number = {10}, 
volume = {112}
}

@article{Lee2017ShellDiffusionMutation, 
year = {2017}, 
title = {{Evidence for improved encapsulated pathway behavior in a bacterial microcompartment through shell protein engineering}}, 
author = {Lee, Marilyn F. Slininger and Jakobson, Christopher M. and Tullman-Ercek, Danielle}, 
journal = {ACS Synthetic Biology}, 
issn = {2161-5063}, 
doi = {10.1021/acssynbio.7b00042}, 
pmid = {28585808}, 
pages = {1880--1891}, 
number = {10}, 
volume = {6}
}

@article{chowdhury_engineering_2019,
	title = {Engineering the {PduT} shell protein to modify the permeability of the 1,2-propanediol microcompartment of Salmonella},
	volume = {165},
	issn = {1350-0872, 1465-2080},
	url = {https://qa.microbiologyresearch.org/content/journal/micro/10.1099/mic.0.000872},
	doi = {10.1099/mic.0.000872},
	pages = {1355--1364},
	number = {12},
	journaltitle = {Microbiology},
	journal = {Microbiology},
	author = {Chowdhury, Chiranjit and Bobik, Thomas A},
	urldate = {2025-04-07},
	date = {2019-12-01},
	year = {2019},
	langid = {english},
}

@article{tasneem_how_2022,
	title = {How pore architecture regulates the function of nanoscale protein compartments},
	volume = {16},
	rights = {https://doi.org/10.15223/policy-029},
	issn = {1936-0851, 1936-086X},
	url = {https://pubs.acs.org/doi/10.1021/acsnano.2c02178},
	doi = {10.1021/acsnano.2c02178},
	pages = {8540--8556},
	number = {6},
	journaltitle = {{ACS} Nano},
	journal = {{ACS} Nano},
	shortjournal = {{ACS} Nano},
	author = {Tasneem, Nuren and Szyszka, Taylor N. and Jenner, Eric N. and Lau, Yu Heng},
	urldate = {2025-04-07},
	date = {2022-06-28},
	year = {2022},
	langid = {english},
}

@article{khmelinskaia_control_2018,
	title = {Control of membrane binding and diffusion of cholesteryl-modified {DNA} origami nanostructures by {DNA} spacers},
	volume = {34},
	issn = {0743-7463, 1520-5827},
	url = {https://pubs.acs.org/doi/10.1021/acs.langmuir.8b01850},
	doi = {10.1021/acs.langmuir.8b01850},
	pages = {14921--14931},
	number = {49},
	journaltitle = {Langmuir},
	journal = {Langmuir},
	shortjournal = {Langmuir},
	author = {Khmelinskaia, Alena and Mücksch, Jonas and Petrov, Eugene P. and Franquelim, Henri G. and Schwille, Petra},
	urldate = {2025-04-27},
	date = {2018-12-11},
	year = {2018},
	langid = {english},
}

@article{rosierProximityinducedCaspase9Activation2020,
  title = {Proximity-induced Caspase-9 activation on a {{DNA}} origami-based synthetic apoptosome},
  author = {Rosier, Bas J. H. M. and Markvoort, Albert J. and Gum{\'i} Audenis, Berta and Roodhuizen, Job A. L. and Den Hamer, Anniek and Brunsveld, Luc and De Greef, Tom F. A.},
  year = {2020},
  month = jan,
  journal = {Nature Catalysis},
  volume = {3},
  number = {3},
  pages = {295--306},
  issn = {2520-1158},
  doi = {10.1038/s41929-019-0403-7},
  url = {https://www.nature.com/articles/s41929-019-0403-7},
  urldate = {2025-06-12},
  langid = {english},
}

@article{Leonard2023MembranesChangeReactionKinetics, 
year = {2023}, 
title = {{The membrane surface as a platform that organizes cellular and biochemical processes}}, 
author = {Leonard, Thomas A. and Loose, Martin and Martens, Sascha}, 
journal = {Developmental Cell}, 
issn = {1534-5807}, 
doi = {10.1016/j.devcel.2023.06.001}, 
pmid = {37419118}, 
pages = {1315--1332}, 
number = {15}, 
volume = {58}
}

@article{Xing.2018, 
year = {2018}, 
title = {{Structures of chaperone-substrate complexes docked onto the export gate in a type III secretion system}}, 
author = {Xing, Qiong and Shi, Ke and Portaliou, Athina and Rossi, Paolo and Economou, Anastassios and Kalodimos, Charalampos G.}, 
journal = {Nature Communications}, 
doi = {10.1038/s41467-018-04137-4}, 
pmid = {29720631}, 
pmcid = {PMC5932034}, 
pages = {1773}, 
number = {1}, 
volume = {9}
}

@article{Phan.2018, 
year = {2018}, 
title = {{Bacterial secretion chaperones: the mycobacterial type VII case}}, 
author = {Phan, Trang H and Houben, Edith N G}, 
journal = {FEMS Microbiology Letters}, 
issn = {0378-1097}, 
doi = {10.1093/femsle/fny197}, 
pmid = {30085058}, 
pmcid = {PMC6109436}, 
pages = {fny197}, 
number = {18}, 
volume = {365}
}

@article{Lang.2026, 
year = {2026}, 
title = {{Oligomerization and positive feedback on membrane binding stabilize PAR-3 asymmetries in the C. elegans zygote}}, 
author = {Lang, Charles F. and Maxian, Ondrej and Anneken, Alexander and Munro, Edwin}, 
journal = {Current Biology}, 
issn = {0960-9822}, 
doi = {10.1016/j.cub.2026.02.036}, 
pmid = {41875854}, 
pages = {1509--1524.e5}, 
number = {6}, 
volume = {36}
}

@article{Hu.2002, 
year = {2002}, 
title = {{Dynamic assembly of MinD on phospholipid vesicles regulated by ATP and MinE}}, 
author = {Hu, Zonglin and Gogol, Edward P. and Lutkenhaus, Joe}, 
journal = {Proceedings of the National Academy of Sciences}, 
issn = {0027-8424}, 
doi = {10.1073/pnas.102059099}, 
pmid = {11983867}, 
pmcid = {PMC124476}, 
pages = {6761--6766}, 
number = {10}, 
volume = {99}
}

@article{rackauckas2017differentialequations,
  title={Differentialequations.jl--a performant and feature-rich ecosystem for solving differential equations in {Julia}},
  author={Rackauckas, Christopher and Nie, Qing},
  journal={Journal of Open Research Software},
  volume={5},
  number={1},
  pages={15},
  year={2017},
  publisher={Ubiquity Press},
  doi={10.5334/jors.151}
}

@article{efronNonparametricEstimatesStandard1981,
  title = {Nonparametric estimates of standard error: {{The}} jackknife, the Bootstrap and other methods},
  shorttitle = {Nonparametric Estimates of Standard Error},
  author = {Efron, Bradley},
  year = 1981,
  journal = {Biometrika},
  volume = {68},
  number = {3},
  pages = {589--599},
  issn = {0006-3444, 1464-3510},
  doi = {10.1093/biomet/68.3.589},
  urldate = {2026-07-01},
  langid = {english}
}

@misc{juliangehringJuliangehringBootstrapjlBootstrap2023,
  title = {Juliangehring/{{Bootstrap}}.Jl: {{Bootstrap}} v2.4.0},
  shorttitle = {Juliangehring/{{Bootstrap}}.Jl},
  author = {Julian Gehring and David Widmann and Dave Kleinschmidt and Rory Finnegan and Bogumi{\l} Kami{\'n}ski and Colin Bowers and Hector Perez and Mike Molignano and {Milan Bouchet-Valat} and Patrick Kofod Mogensen and Tony Kelman and Nikos Ignatiadis},
  year = 2023,
  month = nov,
  doi = {10.5281/ZENODO.10073439},
  urldate = {2026-07-01},
  copyright = {Creative Commons Attribution 4.0 International},
  howpublished = {Zenodo}
}

@misc{zenodo,
  author       = {Swiderski, Richard and
                  Angerpointner, Severin and
                  Frey, Erwin},
  title        = {Code for ``{S}tochastic Yield Catastrophe in Delay-Facilitated Self-Assembly'''},
  month        = jul,
  year         = 2026,
  publisher    = {Zenodo},
  doi          = {10.5281/zenodo.21238865},
  url          = {https://doi.org/10.5281/zenodo.21238865},
  copyright = {Creative Commons Attribution 4.0 International},
  howpublished = {Zenodo}
  }

\end{document}